\documentclass[useAMS, usenatbib]{mn2e}
\usepackage{color}
\usepackage{epsfig}
\usepackage{epstopdf}
\usepackage{amsmath, amssymb}
\usepackage{hyperref}
\usepackage{wasysym} 
\usepackage{subfigure}

\newlength{\plotwidth}
\newlength{\fullwidth}
\renewcommand{\topmargin}{-1cm} 
 
\title[6\lowercase{d}FGS: $z \approx 0$ measurements of $f\sigma_8$ and $\sigma_8$]{The 6\lowercase{d}F Galaxy Survey: $z \approx 0$ measurements of the growth rate and $\sigma_8$}
\author[Beutler et al.]
{\parbox{\textwidth}{Florian Beutler$^{1}$\thanks{E-mail: \texttt{florian.beutler@icrar.org}},
Chris Blake$^2$, Matthew Colless$^3$, D. Heath Jones$^4$,\\
Lister Staveley-Smith$^{1,5}$, Gregory B. Poole$^2$, Lachlan Campbell$^6$, Quentin Parker$^{3,7}$, Will Saunders$^3$, Fred Watson$^3$}\vspace{0.4cm}\\
\parbox{\textwidth}{
$^{1}$International Centre for Radio Astronomy Research (ICRAR), University of Western Australia, 35 Stirling Highway, Crawley WA 6009, Australia\\
$^{2}$Centre for Astrophysics \& Supercomputing, Swinburne University of Technology, P.O. Box 218, Hawthorn, VIC 3122, Australia\\
$^{3}$Australian Astronomical Observatory, PO Box 296, Epping NSW 1710, Australia\\
$^{4}$School of Physics, Monash University, Clayton, VIC 3800, Australia\\
$^{5}$ARC Centre of Excellence for All-sky Astrophysics (CAASTRO)\\
$^{6}$Western Kentucky University, Bowling Green, KY 42101, USA\\
$^{7}$Department of Physics and Astronomy, Faculty of Sciences, Macquarie University, NSW 2109, Sydney, Australia}}

\begin{document}

\label{firstpage}

\maketitle

\begin{abstract}
We present a detailed analysis of redshift-space distortions in the two-point correlation function of the 6dF Galaxy Survey (6dFGS). The $K$-band selected sub-sample which we employ in this study contains $81\,971$ galaxies distributed over $17\,000\,$deg$^2$ with an effective redshift $z_{\rm eff} = 0.067$. By modelling the 2D galaxy correlation function, $\xi(r_p,\pi)$, we measure the parameter combination $f(z_{\rm eff})\sigma_8(z_{\rm eff}) = 0.423 \pm 0.055$, where $f \simeq \Omega_m^{\gamma}(z)$ is the growth rate of cosmic structure and $\sigma_8$ is the r.m.s. of matter fluctuations in $8h^{-1}\,$Mpc spheres.\\
Alternatively, by assuming standard gravity we can break the degeneracy between $\sigma_8$ and the galaxy bias parameter, $b$.  Combining our data with the Hubble constant prior from~\citet{Riess:2011yx}, we measure $\sigma_8 = 0.76 \pm 0.11$ and $\Omega_m = 0.250 \pm 0.022$, consistent with constraints from other galaxy surveys and the Cosmic Microwave Background data from WMAP7.\\
Combining our measurement of $f \sigma_8$ with WMAP7 allows us to test the cosmic growth history and the relationship between matter and gravity on cosmic scales by constraining the growth index of density fluctuations, $\gamma$. Using only 6dFGS and WMAP7 data we find $\gamma = 0.547 \pm 0.088$, consistent with the prediction of General Relativity. We note that because of the low effective redshift of 6dFGS our measurement of the growth rate is independent of the fiducial cosmological model (Alcock-Paczynski effect).  We also show that our conclusions are not sensitive to the model adopted for non-linear redshift-space distortions.\\
Using a Fisher matrix analysis we report predictions for constraints on $f\sigma_8$ for the WALLABY survey and the proposed TAIPAN survey. The WALLABY survey will be able to measure $f\sigma_8$ with a precision of $4-10\%$, depending on the modelling of non-linear structure formation. This is comparable to the predicted precision for the best redshift bins of the Baryon Oscillation Spectroscopic Survey (BOSS), demonstrating that low-redshift surveys have a significant role to play in future tests of dark energy and modified gravity.
\end{abstract}


\section{Introduction}

The distribution of matter in the cosmos depends on the gravitational interaction and the expansion history of the Universe. Assuming that galaxies trace the mass distribution, a measurement of galaxy clustering can be used to derive fundamental properties of the Universe. 

On large scales the movement of galaxies is dominated by the Hubble recession, while on small scales the gravitational field introduces so-called peculiar velocities. Individual galaxy redshifts combine both Hubble recession and peculiar velocities indistinguishably. However, these effects can be statistically distinguished in a large sample of galaxy redshifts. This is the purpose of this paper. The difference between the redshift-inferred distance and the true distance is known as redshift-space distortion. Redshift-space distortions effectively couple the density and velocity fields, complicating the models needed to accurately describe observed galaxy samples. On the other hand they permit measurements of the properties of the galaxy velocity field, which are difficult to access otherwise. In standard gravity we can use the amplitude of peculiar velocities to measure parameters that describe the matter content of the Universe such as $\Omega_m$ and $\sigma_8$. 

In this paper we report measurements of the parameter combination $f(z_{\rm eff})\sigma_8(z_{\rm eff})$, where $f = d\ln(D)/d\ln(a)$ is the growth rate of cosmic structure (in terms of the linear growth factor $D$ and cosmic scale factor $a$) and $\sigma_8$ is the r.m.s. of the matter fluctuations in spheres of $8h^{-1}\,$Mpc. The measurement of $f(z_{\rm eff})\sigma_8(z_{\rm eff})$ can be used to test theories of dark energy and modified gravity, since a stronger gravitational interaction causes a larger growth rate $f$. It is interesting to note in this context the fact that a different form of gravitational interaction on large scales could be responsible for the accelerating expansion of the Universe (e.g.~\citealt{Dvali:2000hr, Wang:2007ht}). Probes such as type Ia supernovae, the Cosmic Microwave Background (CMB) or Baryon Acoustic Oscillations, which have proven the existence of the current acceleration of the expansion of the Universe (e.g.~\citealt{Blake:2011en}), cannot distinguish between acceleration due to a dark energy component with negative pressure or due to a modification of General Relativity.  However, measurements of the growth of structure are able to distinguish between these models.

In order to measure $f(z_{\rm eff})\sigma_8(z_{\rm eff})$ we have to model the effect of redshift-space distortions on the correlation function. While on large scales linear theory can be used to model these effects, on smaller scales non-linear contributions complicate the process. 
Several new approaches have been suggested in recent years to extend linear theory. We will discuss some of these models and apply them to our dataset.

Redshift-space distortions have previously been analysed using both the correlation function and power spectrum using data from the 2dF Galaxy Redshift Survey (2dFGRS;~\citealt{Peacock:2001gs, Hawkins:2002sg, Cole:2005sx}) and the Sloan Digital Sky Survey (SDSS;~\citealt{Tegmark:2003ud,Zehavi:2004zn,Tegmark:2006az,Cabre:2008sz,Song:2010kq,Samushia:2011cs}). More recently it has become possible to do similar studies at higher redshift using the VVDS~\citep{Guzzo:2008ac}, WiggleZ~\citep{Blake:2011rj} and VLT VIMOS surveys~\citep{Bielby:2010ps}. 

Redshift-space distortion measurements are also sensitive to the overall amplitude of the clustering pattern of matter, commonly parameterised by $\sigma_8$~\citep{Lahav:2001sg}. This parameter is used to normalise the amplitude of clustering statistics such as the correlation function, $\xi \propto \sigma_8^2$. From the CMB we have a very accurate measurement of the matter fluctuations in the early universe (the scalar amplitude $A_s$) at the time of decoupling, $z_*$. In order to derive $\sigma_8(z{=}0)$ we have to extrapolate this measurement to redshift zero, involving assumptions about the expansion history of the Universe. The CMB constraint on $\sigma_8$ heavily depends on these assumptions. Hence there is a clear advantage in obtaining low-redshift measurements of this parameter. The 6dF Galaxy Survey, which we analyse in this study, is one of the largest galaxy redshift surveys available. Its very small effective redshift and wide areal coverage ($41\%$ of the sky) make it a powerful sample for the study of the local galaxy distribution. 

Galaxy surveys usually have to consider degeneracies between redshift-space distortions and the Alcock-Paczynski effect, which arises from the need to assume a cosmological model to transform redshifts into distances. At low redshift this effect is very small, meaning that our measurement is fairly independent of the choice of the fiducial cosmological model.

While at high redshift ($z > 1$) the matter density dominates both the expansion of the Universe and the growth of perturbations, at low redshift these two are partially decoupled, with dark energy mostly dominating the background expansion and the matter density dominating the growth of perturbations. As a result in $\Lambda$CDM and most proposed modified gravity models, low redshift measurements of the growth rate have a better constraining power than high redshift measurements. The measurement of the growth rate in 6dFGS therefore not only provides a new independent data point at very low redshift, but also promises to make a valuable contribution to tests of General Relativity on cosmic scales.

The outline of this paper is as follows: In section~\ref{sec:survey} we introduce the 6dF Galaxy Survey. In section~\ref{sec:data} we describe the details of the correlation function estimate and introduce the 2D correlation function of 6dFGS. In section~\ref{sec:error} we derive the covariance matrix for the 2D correlation function. In section~\ref{sec:theory} we summarise the theory of redshift-space distortions, including extensions to the standard linear approach. We also discuss wide-angle effects and other systematics, such as the Alcock-Paczynski effect. In section~\ref{sec:fit} we fit the 2D correlation function to derive $g_{\theta}(z_{\rm eff})=f(z_{\rm eff})\sigma_8(z_{\rm eff})$ and $\sigma_8$. Cosmological implications are investigated in section~\ref{sec:impl}. In section~\ref{sec:future} we make Fisher matrix predictions for two future low redshift galaxy surveys, WALLABY and the proposed TAIPAN survey. We conclude in section~\ref{sec:conc}.

Throughout the paper we use $r$ to denote real space separations and $s$ to denote separations in redshift-space. Our fiducial model assumes a flat universe with $\Omega^{\rm fid}_m = 0.27$, $w^{\rm fid} = -1$ and $\Omega_k^{\rm fid} = 0$. The Hubble constant is set to $H_0 = 100h\,$km s$^{-1}$Mpc$^{-1}$.

\section{The 6\lowercase{d}F Galaxy Survey}
\label{sec:survey}

The 6dF Galaxy Survey~\citep[6dFGS;][]{Jones:2004zy,Jones:2006xy,Jones:2009yz} is a near-infrared selected ($JHK$) redshift survey of $125\,000$ galaxies across four-fifths of the southern sky, with secondary samples selected in $b_{\rm J}$ and $r_{\rm F}$. The $|b| < 10^\circ$ region around the Galactic Plane is avoided by the $JHK$ surveys to minimise Galactic extinction and foreground source confusion in the Plane (as is $|b| < 20^\circ$ for $b_{\rm J}$ and $r_{\rm F}$). The near-infrared photometric selection was based on total magnitudes from the Two-Micron All-Sky Survey -- Extended Source Catalog~\citep[2MASS XSC;][]{Jarrett:2000me}. The spectroscopic redshifts of 6dFGS were obtained with the Six-Degree Field (6dF) multi-object spectrograph of the UK Schmidt Telescope (UKST) between 2001 and 2006. The effective volume of 6dFGS is about the same as the 2dF Galaxy Redshift Survey~\citep[2dFGRS;][]{Colless:2001gk} and is a little under a third that of the Sloan Digital Sky Survey main spectroscopic sample at its Seventh Data Release~\citep[SDSS DR7;][]{Abazajian:2008wr}. A subset of early-type 6dFGS galaxies (approximately $10\,000$) have measured line-widths that will be used to derive Fundamental Plane distances and peculiar motions.

The 6dFGS $K$-selected sample used in this paper contains 81\,971 galaxies selected to a faint limit of $K = 12.75$. The 2MASS magnitudes are on the Vega system. The mean completeness of the 6dFGS is 92 percent and median redshift is $z = 0.05$. Completeness corrections are derived by normalising completeness-apparent magnitude functions so that, when integrated over all magnitudes, they equal the measured total completeness on a particular patch of sky. This procedure is outlined in the luminosity function evaluation of~\citet{Jones:2006xy} and also in Jones et al., (in prep). The original survey papers~\citep{Jones:2004zy,Jones:2009yz} describe in full detail the implementation of the survey and its associated online database.

\begin{figure}
\begin{center}
\epsfig{file=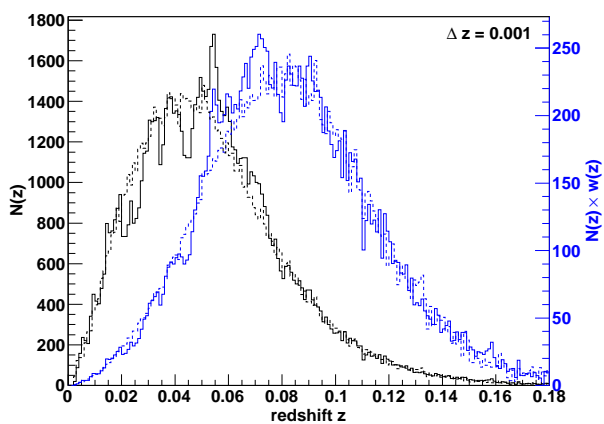,width=8cm}
\caption{The solid black line shows the 6dFGS redshift distribution, while the dashed black line shows one of the random mock catalogues containing the same number of galaxies. The blue solid and dashed lines show the distribution after weighting with $P_0 = 1600h^3\,$Mpc$^{-3}$ (see section~\ref{sec:dweight} for more details on the employed weighting scheme).}
\label{fig:wred}
\end{center}
\end{figure}

The clustering in a galaxy survey is estimated relative to a random (unclustered) distribution which follows the same angular and redshift selection function as the galaxy sample itself. We base our random mock catalogue generation on the 6dFGS luminosity function, where we use random numbers to pick volume-weighted redshifts and luminosity function-weighted absolute magnitudes. We then test whether the redshift-magnitude combination falls within the 6dFGS $K$-band faint and bright apparent magnitude limits ($8.75 \leq K \leq 12.75$).

Figure~\ref{fig:wred} shows the redshift distribution of the 6dFGS $K$-selected sample (black solid line) compared to a mock catalogue with the same number of galaxies (black dashed line).

\section{Correlation function measurement}
\label{sec:data}

We calculate the co-moving distances for each galaxy using the measured redshift
\begin{equation}
D_C = \frac{c}{H_0}\int^z_0\frac{dz'}{E(z')}
\end{equation} 
with
\begin{equation}
E(z) = \left[\Omega^{\rm fid}_m(1+z)^3 + \Omega^{\rm fid}_{\Lambda}\right]^{1/2},
\end{equation}
where we assume a flat universe with $\Omega_k^{\rm fid} = 0$ and $\Omega^{\rm fid}_{\Lambda} = 1 - \Omega^{\rm fid}_m$ and describe dark energy as a cosmological constant ($w^{\rm fid} = -1$). Given the low redshift of our dataset, these assumptions have a very small impact on our final results (see section~\ref{sec:AP}).

We define the positions of two galaxies as $\vec{s_1}$ and $\vec{s_2}$. The redshift-space separation is then given by $\vec{h} = \vec{s}_1 - \vec{s}_2$, while $\vec{s} = (\vec{s}_1 + \vec{s}_2)/2$ is the mean distance to the pair of galaxies. Now we can calculate the separation along the line-of-sight $\pi$ and the separation across the line-of-sight $r_p$
\begin{align}
\pi &= \frac{|\vec{s}\cdot\vec{h}|}{|\vec{s}|},\\
r_p &= \sqrt{|\vec{h}|^2 - \pi^2}.
\end{align}
The absolute separation is then given by $s = \sqrt{\pi^2 + r_p^2}$.

We measure the separation between all galaxy pairs in our survey and count the number of such pairs in each separation bin. We do this for the 6dFGS data catalogue, a random catalogue with the same selection function, and a combination of data-random pairs. We call the pair-separation distributions obtained from this analysis step $DD$, $RR$ and $DR$, respectively. In the analysis we used $30$ random catalogues with the same size as the real data catalogue and average $DR$ and $RR$. The redshift-space correlation function itself is then given by~\cite{Landy:1993yu}:
\begin{equation}
\xi'_{\rm data} = 1 + \frac{DD}{RR} \left(\frac{n_r}{n_d} \right)^2 - 2\frac{DR}{RR} \left(\frac{n_r}{n_d} \right),
\label{eq:LS2}
\end{equation}
where the ratio $n_r/n_d$ is given by
\begin{equation}
\frac{n_r}{n_d} = \frac{\sum^{N_r}_iw_i}{\sum^{N_d}_jw_j}
\end{equation}
and the sums go over all random ($N_r$) and data ($N_d$) galaxies. The galaxies are weighted by the inverse completeness $C_i$ of their area of the sky
\begin{equation}
w_i(z) = C_i.
\label{eq;compl}
\end{equation}
We will discuss further weighting techniques in the next section. 

There is a possible bias in the estimation of the correlation function due to the fact that we estimate both the mean density and the pair counts from the same survey. This leads to a non-zero difference between the true correlation function estimate of an ensemble of surveys and the ensemble average of $\xi(s)$ from each survey. This is commonly known as the integral constraint (e.g.~\citealt{Peebles:1980}), which can be calculated as (see e.g.~\citealt{Roche:2002vj})
\begin{equation}
ic = \frac{\sum \xi_{\rm model}RR}{\sum RR}
\end{equation}
and enters our correlation function estimate as
\begin{equation}
\xi_{\rm data} = \xi'_{\rm data} + ic,
\end{equation}
where $\xi'_{\rm data}$ is the redshift-space correlation function from eq.~\ref{eq:LS2} and $\xi_{\rm model}$ is the model for the correlation function. In 6dFGS $ic$ is typically around $6\times 10^{-4}$ and so has no significant impact on the final result.

\begin{figure}
\begin{center}
\epsfig{file=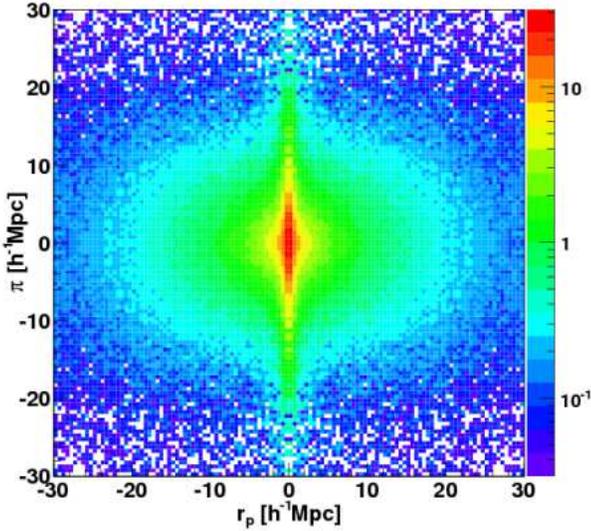,width=8cm}
\caption{The 2D correlation function of 6dFGS using a density weighting with $P_0 = 1600h^3\,$Mpc$^{-3}$. For reasons of presentation we binned the correlation function in $0.5h^{-1}\,$Mpc bins, while in the analysis we use larger bins of $2h^{-1}\,$Mpc. Both redshift-space distortion effects are visible: the ``finger-of-God'' effect at small angular separation $r_p$, and the anisotropic (non-circular) shape of the correlation function at large angular separations.}
\label{fig:kaiser}
\end{center}
\end{figure}

In Figure~\ref{fig:kaiser} we show the 2D correlation function calculated from the 6dFGS dataset. In this Figure we use bins of $0.5h^{-1}\,$Mpc, while for the analysis later on we use larger bins of $2h^{-1}\,$Mpc (see Figure~\ref{fig:kaiser_models}). The figure shows clearly the two effects of redshift-space distortions which we will discuss later in section~\ref{sec:theory}, the ``finger-of-God'' effect at small $r_p$, and the linear infall effect at larger $r_p$ which gives the correlation function a non-circular shape.

\subsection{Density weighting}
\label{sec:dweight}

In Fourier space the error  in measuring the amplitude of a mode of the linear power spectrum\footnote{As the correlation function and power spectrum are related by a Fourier transform, the following discussion also holds true for a correlation function measurement.} is given by
\begin{equation}
\sigma_{P(k)} = (b + f\mu^2)^2P(k) + \langle N \rangle,
\end{equation}
where $b$ is the linear bias, $f$ is the growth rate, $\mu$ is the cosine of the angle to the line of sight and $P(k)$ is the matter power spectrum. The first term on the right hand side of this equation represents the sample-variance error, while the second term ($\langle N \rangle$) represents the Poisson error.

\begin{figure}
\centering

\subfigure[]{
   \epsfig{file=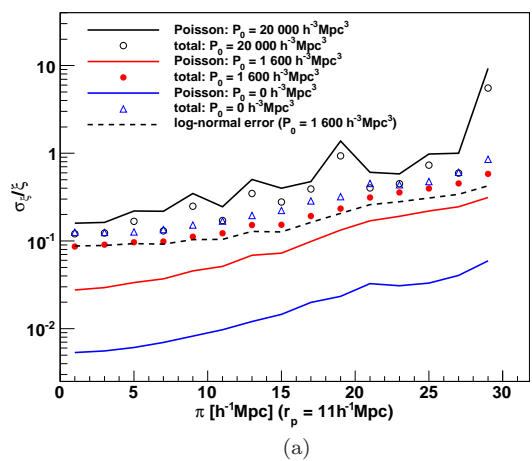,width=8cm}
   \label{fig:kaiser_error}
 }

 \subfigure[]{
   \epsfig{file=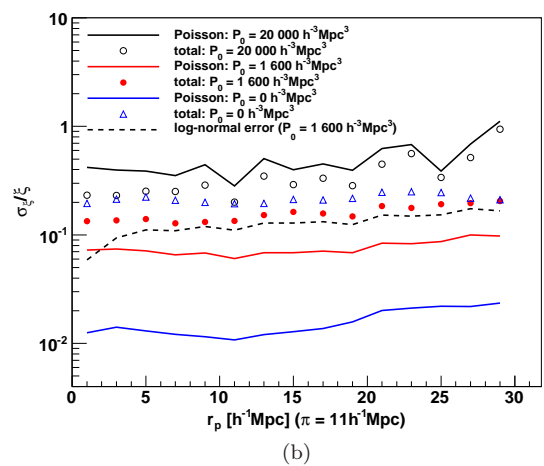,width=8cm}
   \label{fig:kaiser_error2}
 }

\label{fig:kaiser_errors}
\caption{(a) The relative error in the 2D correlation function as a function of line-of-sight separation $\pi$ at a fixed $r_p = 11h^{-1}\,$Mpc. Other regions of the 2D correlation function behave in a similar manner. The solid lines show the Poisson error for different values of $P_0$, while the data points show the total (Poisson + sample variance) error obtained as the diagonal of the covariance matrix derived using jack-knife re-sampling. The purpose of the weighting ($P_0$) is to minimise the total error, which is achieved for a value of $P_0 \approx 1600h^{-3}\,$Mpc$^{3}$. The weighting reduces the error by almost a factor of two on most scales. The dashed line shows the error derived from log-normal realisations using $P_0 = 1600h^{-3}\,$Mpc$^{3}$ and is in very good agreement with the jack-knife error.
(b) same as (a) for a fixed line-of-sight separation $\pi = 11h^{-1}\,$Mpc.}
\end{figure}

If the sample-variance error is dominant we can reduce the power spectrum error by employing a weighting scheme which depends upon the galaxy density $n(z)$, such as the one suggested by~\citet{Feldman:1993ky}
\begin{equation}
w_i(z) = \frac{1}{1 + n(z)P_0},
\label{eq:static}
\end{equation}
where $P_0$ describes the amplitude of the weighting. A stronger weighting (larger value of $P_0$) yields a smaller sample-variance error since it increases the survey volume by up-weighting sparsely sampled regions. However, such a weighting scheme also increases the Poisson error because it shifts the effective redshift to larger values with a smaller galaxy number density. This is illustrated in Figure~\ref{fig:kaiser_error} and~\ref{fig:kaiser_error2}. Such a weighting scheme is standard for large scale structure analyses.

In a magnitude-limited sample such as 6dFGS, up-weighting higher redshift galaxies also has the effect of shifting the galaxy bias to larger values. The sample-variance error is proportional to the clustering amplitude, and so a larger bias results in a larger error. However, the weighting will still ensure that the relative error of the power spectrum, $\sigma_{P(k)}/P(k)$, is minimised. The redshift-space distortion signal is inversely proportional to the galaxy bias, $\beta \simeq \Omega_m^{\gamma}(z)/b$. If weighting increases the bias $b$, it also reduces the signal we are trying to measure. We therefore must investigate whether the advantage of the weighting (the reduced relative error) outweighs the disadvantage (increasing galaxy bias).

The situation is very different for measuring a signal that is proportional to the clustering amplitude, such as the baryon acoustic peak. In this case the error and the signal are proportional to the bias, and so weighting will always be beneficial. We stress that an increasing bias with redshift is expected in almost all galaxy redshift surveys. Therefore redshift-space distortion studies should first test whether galaxy weighting improves the measurement. The 6dF Galaxy Survey is quite sensitive to the weighting scheme employed because it has a high galaxy density, making the sample-variance error by far the dominant source of error.

Finally, we have to consider the correlation between the bins in the measured power spectrum or correlation function. If the error is sample-variance dominated, the bins will show large correlation (especially in the correlation function), while in the case of Poisson-noise dominated errors, the correlation is much smaller. Weighting will always increase the Poisson noise and hence reduce the correlation between bins.

All of the density weighting effects discussed above need to be considered, before deciding which weighting is best suited to the specific analysis. We summarise as follows:
\begin{itemize}
\item If a galaxy sample is sample-variance limited, density weighting will reduce the (relative) error of the clustering measurement.
\item In most galaxy redshift surveys, the density weighting increases the galaxy bias, which reduces the redshift-space distortion signal, by flattening the clustering anisotropy. 
\item Galaxy weighting also reduces the correlation between bins in the power spectrum and correlation function.
\end{itemize}
Whether the signal-to-noise is actually improved by density weighting depends on the specific sample. For the 6dFGS redshift-space distortion analysis we found $P_0 \approx 1600h^{-3}\,$Mpc$^{3}$ leads to the most accurate constraint on the growth rate.

This discussion also indicates that low-biased galaxy samples have an advantage over a highly biased galaxy sample in measuring redshift-space distortions. We will discuss this point further in section~\ref{sec:future}.

\section{Error estimate}
\label{sec:error}

In this section we will derive a covariance matrix for the 2D correlation function using jack-knife re-sampling. We also use log-normal realisations to test the jack-knife covariance matrix.

\subsection{Jack-knife re-sampling}

We divide the dataset into $N = 480$ subsets, selected in R.A. and Dec. Each re-sampling step excludes one subset before calculating the correlation function. The $N-1$ remaining subsets have a volume which is $(N-1)/N$ times the volume of the original data. The covariance matrix is then given by
\begin{equation}
C_{ij} = \frac{(N-1)}{N}\sum^N_{k=1}\left[\xi^k(s_i) - \overline{\xi}(s_i)\right]\left[\xi^k(s_j) - \overline{\xi}(s_j)\right],
\end{equation}
where $\xi^k(s_i)$ is the correlation function estimate at separation $s_i$ with the exclusion of subset $k$. The mean value is defined as
\begin{equation}
\overline{\xi}(s_i) = \frac{1}{N}\sum^N_{k=1}\xi^k(s_i).
\end{equation}
The case $i = j$ gives the error ignoring correlations between bins $\sigma_i^2 = C_{ii}$. 

\subsection{Log-normal realisations}

We can create a log-normal realisation~\citep{Coles:1991if,Cole:2005sx, Percival:2006gt, Blake:2011rj, Beutler:2011hx} of a galaxy survey by deriving a density field from a model power spectrum, $P(k)$, assuming Gaussian fluctuations. This density field is then Poisson sampled, taking into account the window function and the total number of galaxies. The assumption that the input power spectrum has Gaussian fluctuations can only be used if the fluctuations are much smaller than the mean density, otherwise the Gaussian model assigns a non-zero probability to regions of negative density. A log-normal random field, $LN(\vec{x})$, can avoid this unphysical behaviour. It is obtained from a Gaussian field $G(\vec{x})$ by
\begin{equation}
LN(\vec{x}) = \exp[G(\vec{x})] - 1,
\end{equation}
which is  positive-definite but approaches $G(\vec{x})$ whenever the perturbations are small. As an input power spectrum for the log-normal field we use the linear model~\citep{Kaiser:1987qv}, given by
\begin{equation}
P_g(k,\mu) = b^2(1 + \beta\mu^2)^2P_{\rm \delta\delta}(k)
\end{equation}
with $b = 1.47$ and $\beta = 0.35$, which is consistent with the parameters we measure for 6dFGS (see section~\ref{sec:fit}). $P_{\rm \delta\delta}(k)$ is a linear density power spectrum in real space obtained from CAMB~\citep{Lewis:1999bs} and $P_g(k,\mu)$ is the galaxy power spectrum in redshift-space. Our method is explained in more detail in~\citet{Beutler:2011hx}, appendix A.

We produce $N = 1500$ such realisations and calculate the 2D correlation function for each of them, deriving a covariance matrix
\begin{equation}
\begin{split}
C_{ij} = \frac{1}{N-1}&\sum^N_{n=1}\left[\xi^i_n(r_p,\pi) - \overline{\xi^i}(r_p,\pi)\right]\cr
&\times\left[\xi^j_n(r_p,\pi) - \overline{\xi^j}(r_p,\pi)\right],
\end{split}
\end{equation}
where the mean value $\overline{\xi^i}(r_p,\pi)$ is defined as
\begin{equation}
\overline{\xi^i}(r_p,\pi) = \frac{1}{N}\sum^N_{n=1}\xi^i_n(r_p,\pi)
\end{equation}
and $\xi^i_n(r_p,\pi)$ is the 2D correlation function estimate of realisation $n$ at a specific separation $(r_p, \pi)$.

\subsection{Discussion: Error analysis} 

\begin{figure}
\begin{center}
   \epsfig{file=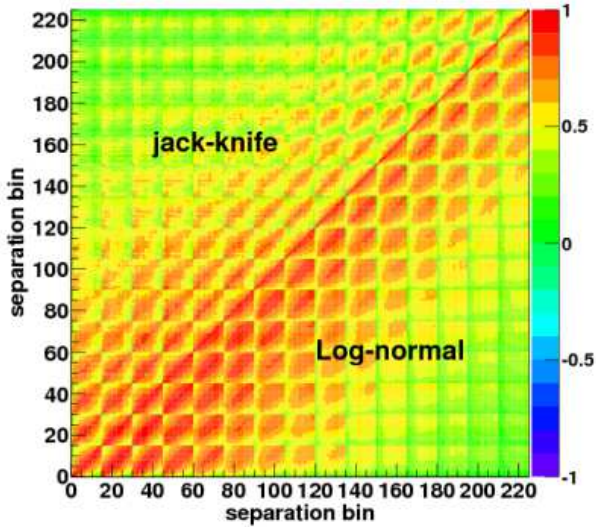,width=8cm}
\caption{The correlation matrix for the 2D correlation function $\xi(r_p,\pi)$ with a bin size of $2\times 2h^{-1}\,$Mpc. The upper-left corner shows the jack-knife estimate, while the lower-right corner shows the result of using $1500$ log-normal realisations. Since this plot shows the correlation of all $15\times 15$ bins it contains $225\times 225$ entries.}
   \label{fig:cov2D}
\end{center}
\end{figure}

\begin{figure}
\begin{center}
   \epsfig{file=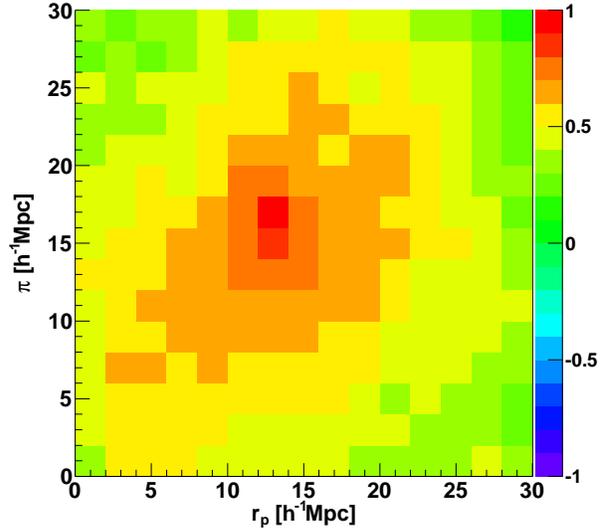,width=8cm}
\caption{This plot shows the correlation of bin $127$ ($r_p = 13h^{-1}\,$Mpc, $\pi = 17h^{-1}\,$Mpc) with all other bins in the 2D correlation function, derived using jack-knife re-sampling. It corresponds to row/column $127$ of the jack-knife correlation matrix which is shown in Figure~\ref{fig:cov2D} (upper-left corner).}
   \label{fig:cov2D_2}
\end{center}
\end{figure}

Figure~\ref{fig:cov2D} shows the correlation matrix
\begin{equation}
r_{ij} = \frac{C_{ij}}{\sqrt{C_{ii}C_{jj}}},
\end{equation}
derived from the two different covariance matrices, $C_{ij}$. We plot the jack-knife result in the upper-left corner and the log-normal result in the lower-right corner. Both the correlation and covariance matrix are symmetric. While the log-normal correlation matrix indicates somewhat more correlation between bins, overall the correlation matrices are in rough agreement. The diagonal errors for both the log-normal and jack-knife covariance matrices are plotted in Figures~\ref{fig:kaiser_error} and \ref{fig:kaiser_error2} and show very good agreement at large scales.

Every row/column in Figure~\ref{fig:cov2D} shows the correlation of one bin with all other bins in $\xi(r_p, \pi)$. Figure~\ref{fig:cov2D_2} shows an example of such a row/column obtained from jack-knife re-sampling as a $15\times 15$ matrix, in this case for bin $127$ ($r_p = 13h^{-1}\,$Mpc, $\pi = 17h^{-1}\,$Mpc).

Log-normal realisations do not account for non-linear mode coupling and are very model-dependent in the quasi-linear and non-linear regime. 
Since our analysis relies on fits to fairly small scales, we decided to use the jack-knife covariance matrix in our analysis. However, we find that none of the results reported in this paper depend significantly on which of the two covariance matrices is used.

\section{Modelling the 2D correlation function}
\label{sec:theory}

In this section we discuss the theory of redshift-space distortions, starting with the standard linear perturbation theory. We then describe different approaches for extending the linear model to include non-linear structure formation. We also discuss deviations from the plane-parallel approximation.

\subsection{Linear redshift-space distortions}

The position of a galaxy in real space $\vec{r} = (x,y,z)$ is mapped to the position in redshift-space, $\vec{s}$, via
\begin{equation}
\vec{s} = \vec{r} + \frac{v_z(\vec{r})}{aH(z)}\hat{z},
\end{equation}
where the unit vector $\hat{z}$ indicates the line-of-sight direction, and the quantity $v_z$ is the line-of-sight component of the velocity field, namely $v_z = \vec{v}\cdot\hat{z}$. The scale factor $a$ is defined as $1/(1+z)$ and $H(z)$ is the Hubble constant at redshift $z$. The second term in the equation above represents peculiar velocities caused by gravitational interaction. 
On small scales this elongates structures along the line-of-sight and leads to the so-called ``finger-of-God'' effect~\citep{Jackson:2008yv}. On large scales matter falls in towards over-dense regions and systematically influences our distance measurement, making the over-densities appear more over-dense. The latter effect can be described by linear theory, while the ``finger-of-God'' effect is a non-linear phenomenon.

The model of linear redshift-space distortions has been developed by~\citet{Kaiser:1987qv} and~\citet{Hamilton:1992} assuming a plane-parallel approximation. In Fourier space the linear model can be written as 
\begin{equation}
P_g(k,\mu) = b^2(1 + \beta\mu^2)^2P_{\delta\delta}(k),
\end{equation}
where $P_{\delta\delta}(k)$ is the matter density power spectrum and $P_g(k,\mu)$ is the galaxy density power spectrum. In this model, linear redshift-space distortions are quantified by the parameter $\beta$, which is defined as
\begin{equation}
\beta = \frac{1}{b}\frac{d\ln D(z)}{d\ln(a)} \simeq \frac{\Omega_m^{\gamma}(z)}{b},
\end{equation} 
where $b$ is the linear galaxy bias factor, $D(z)$ is the growth factor and $\gamma$ is the gravitational growth index, which takes the value $\gamma=0.55$ in $\Lambda$CDM~\citep{Linder:2005in}.
$\Omega_m(z)$ is the matter density at redshift $z$, and is defined as 
\begin{equation}
\Omega_m(z) = \frac{H_0^2}{H(z)^2}\Omega_m(z=0)(1+z)^{3}
\end{equation}
with
\begin{equation}
\frac{H^2_0}{H(z)^2} = \left[\Omega_m(1+z)^{3}  + \Omega_{\Lambda}\right]^{-1},
\end{equation}
where we again follow our fiducial model of $\Omega^{\rm fid}_k = 0$ and $w^{\rm fid} = -1$. Modifications of the gravitational force mainly affect $\gamma$, while changes in the expansion history of the Universe affect $\Omega_m(z)$.

\subsection{Parameterisation}

From now on we will formulate our equations in terms of $g_{\theta}(z) = f(z)\sigma_8(z)$ and $g_b(z) = b\sigma_8(z)$. We choose the parameter set [$g_{\theta}$, $g_b$] instead of [$\beta$, $\sigma_8$, $b$], because $\sigma_8$ and the linear bias, $b$, are degenerate and difficult to disentangle~\citep{White:2008jy,Song:2008qt, Song:2010kq}. Within this parameterisation the power spectrum is expressed as 
\begin{equation}
P_g(k) = b^2 P_{\delta\delta}(k) = g^2_{b}Q_{\delta\delta}(k),
\end{equation}
where $Q_{\delta\delta}(k)$ is the unnormalised matter density power spectrum (see eq.~\ref{eq:D3}).  In terms of our new parameters, we can write
\begin{equation}
\beta = \frac{g_{\theta}}{g_{b}}.
\end{equation}

\subsection{Extensions to linear theory}
\label{sec:model}

For the remainder of this section we will discuss possible extensions of the linear model to include non-linear structure formation and wide-angle effects. The simplest model of non-linearities is the so-called streaming model~\citep{Peebles:1980,Hatton:1997xs}. Here, the redshift-space correlation function is just a convolution of the linear correlation function in redshift-space with a pairwise velocity probability density function, $F(v)$,
\begin{equation}
\xi_{\rm st}(r_p,\pi) = \int^{\infty}_{-\infty} \xi\left(r_p, \pi - \frac{v}{H(z_{\rm eff})a_{\rm eff}}\right)F(v)dv,
\label{eq:stream}
\end{equation}
where $a_{\rm eff} = 1/(1+z_{\rm eff})$ is the scale factor and $H(z_{\rm eff})$ is the Hubble constant at the effective redshift. We chose $F(v)$ to be an exponential distribution~\citep{Peacock:1996ci}
\begin{equation}
F(v) = \frac{1}{\sigma_p\sqrt{2}}\exp\left[\frac{-\sqrt{2}|v|}{\sigma_p}\right],
\label{eq:fv}
\end{equation}
which has been shown to successfully describe observations~\citep{Davis:1982gc,Fisher:1993ye,Marzke:1995yk,Landy:2002xg}. Although the pairwise velocity dispersion within a halo is expected to follow a Gaussian distribution instead of an exponential, galaxies populate halos of a wide range of masses and velocity dispersions, which combine to approximately form an exponential function~\citep{Sheth:1995is,Diaferio:1996de,Seto:1997mv}. The parameter $\sigma_p$ depends on galaxy type (e.g.~\citealt{Madgwick:2003bd}) and hence its use for cosmological constraints is limited. 

In recent years many improvements to the model discussed above have been suggested. We will initially discuss these models in Fourier space because the theoretical motivation is clearer and many expressions needed for these models simplify considerably. However, since our data is in real space we will also give real-space expressions later on. We start with the streaming model of eq.~\ref{eq:stream}, which in Fourier space is given by
\begin{equation}
P_g(k,\mu) = b^2(1 + \beta\mu^2)^2P_{\delta\delta}(k)\frac{1}{1 + k^2\mu^2\sigma_p^2/2}.
\label{eq:nl1}
\end{equation}
Hence in Fourier space, the convolution with an exponential function becomes a multiplication by a Lorentzian distribution.

The model above assumes that there is a perfect correlation between the velocity field and the density field, which is given by $P_{\delta\delta}(k) = P_{\delta\theta}(k) = P_{\theta\theta}(k)$ where $P_{\delta\delta}(k)$ is the matter density power spectrum as before. $P_{\theta\theta}(k) = \langle|\theta_k|^2\rangle$ is the velocity divergence power spectrum where $\theta =  \vec{\nabla} \cdot \vec{v}$ is the velocity divergence, and $P_{\delta\theta}(k)$ is the cross power spectrum. Non-linear effects will violate these assumptions, since the density power spectrum is expected to increase in amplitude at small scales because of non-linear effects, while the velocity field becomes randomised at small scales (e.g. within virialized galaxy clusters) and hence $P_{\theta\theta}(k)$ will decrease in amplitude (e.g.~\citealt{Carlson:2009it}). \citet{Scoccimarro:2004tg} suggested expressing the 2D power spectrum without the assumption of linear relations between the density field and velocity field, by
\begin{equation}
\begin{split}
P_g(k,\mu) &= F_q(k,\mu,\sigma_v)\cr
&\times\left[b^2P_{\delta\delta}(k) + 2\mu^2bfP_{\delta\theta}(k) + \mu^4f^2P_{\theta\theta}(k)\right]\cr
&= F_q(k,\mu,\sigma_v)\cr
&\times\left[g_b^2Q_{\delta\delta}(k) + 2\mu^2g_bg_{\theta}Q_{\delta\theta}(k) + \mu^4g_{\theta}^2Q_{\theta\theta}(k)\right],
\end{split}
\label{eq:nl4}
\end{equation}
where the different $Q_{xy}$ are defined as 
\begin{equation}
\begin{split}
Q_{\delta\delta}(k) &= P_{\delta\delta}(k)/\sigma_8(z_{\rm eff})^2,\\
Q_{\delta\theta}(k) &= P_{\delta\theta}(k)/\sigma_8(z_{\rm eff})^2,\\
Q_{\theta\theta}(k) &= P_{\theta\theta}(k)/\sigma_8(z_{\rm eff})^2.
\end{split}
\label{eq:D3}
\end{equation}
and the damping function $F_q(k,\mu,\sigma_v)$ is usually chosen to be a Gaussian of the form 
\begin{equation}
F_q(k,\mu,\sigma_v) = e^{-(k\mu\sigma_v)^2}.
\end{equation}
The parameter $\sigma_v$ quantifies the non-linear dispersion in the bulk motion of halos. It is different to the $\sigma_p$ parameter we introduced earlier that describes small-scale randomised motion (e.g. of galaxies within a halo). We can derive $\sigma_v$ from the velocity power spectrum as
\begin{equation}
\sigma_v^2(z) = \frac{g_{\theta}(z)^2}{6\pi^2}\int^{\infty}_{0}Q_{\theta\theta}(k)dk.
\label{eq:sigv}
\end{equation}
\citet{Jennings:2010uv} provide fitting formulae for $P_{\delta\theta}(k)$ and $P_{\theta\theta}(k)$ derived from N-body simulations. They find the following relation between the different power spectra
\begin{equation}
P_{xy}(k) = \frac{\alpha_0\sqrt{P_{\delta\delta}(k)} + \alpha_1P^2_{\delta\delta}(k)}{\alpha_2 + \alpha_3P_{\delta\delta}(k)},
\label{eq:nl5}
\end{equation}
where $P_{\delta\delta}(k)$ can be obtained from CAMB by including halofit~\citep{Smith:2002dz}. For the cross power spectrum $P_{xy}(k) = P_{\delta\theta}(k)$ we use the (updated) parameters ($\alpha_0$, $\alpha_1$, $\alpha_2$, $\alpha_3$) = ($-12\,483.8$, $2.55430$, $1\,381.29$, $2.54069$) and for $P_{xy}(k) = P_{\theta\theta}(k)$ we use ($-12\,480.5$, $1.52404$, $2\,165.87$, $1.79640$). The fitting formula reproduces $P_{\theta\theta}$ to better than $1\%$ for $k<0.4h\,$Mpc$^{-1}$, to $10\%$ for $0.4 < k < 0.7h\,$Mpc$^{-1}$, and to $15\%$ for $0.7 < k < 1h\,$Mpc$^{-1}$. It also reproduces $P_{\delta\theta}$ to less than $4\%$ over the whole range $k < 1 h\,$Mpc$^{-1}$ (Jennings, private communication). We cut off the integral when the Jennings formula predicts negative values, although because of the high precision of the fitting formula up to large $k$, such a cut-off will not affect our measurement.

We can express~eq.~\ref{eq:nl4} in real-space as
\begin{equation}
\begin{split}
\xi_{\rm Sc}(r_p,\pi) &= \left[g_b^2\xi_{0,\delta\delta}(r) + \frac{2}{3}g_bg_{\theta}\xi_{0,\delta\theta}(r) + \frac{1}{5}g^2_{\theta}\xi_{0,\theta\theta}(r)\right]\mathcal{P}_0(\mu)\cr
&+\left[\frac{4}{3}g_bg_{\theta}\xi_{2,\delta\theta}(r) + \frac{4}{7}g^2_{\theta}\xi_{2,\theta\theta}(r)\right]\mathcal{P}_2(\mu)\cr
&+ \frac{8}{35}g^2_{\theta}\xi_{4,\theta\theta}(r)\mathcal{P}_4(\mu),
\end{split}
\label{eq:song}
\end{equation}
where $\mathcal{P}_{\ell}(\mu)$ are the Legendre polynomials and the spherical harmonic moments $\xi_{\ell,xy}(r)$ are given by
\begin{equation}
\begin{split}
\xi_{\ell,xy}(r) &= \int^{\infty}_{0}\int^1_{-1}\frac{k^2dkd\mu}{(2\pi)^2}e^{-(k\mu\sigma_v)^2}\cr
&\times \cos(kr\mu)Q_{xy}(k)\mathcal{P}_{\ell}(\mu).
\end{split}
\label{eq:xilm1}
\end{equation}
Appendix~\ref{sec:ana} shows a partial analytic solution for the double integral above.

We therefore have two different models which we will apply to our data: The simple streaming model $\xi_{\rm st}(r_p,\pi)$ and the Scoccimarro models $\xi_{\rm Sc}(r_p,\pi)$. Note that $\xi_{\rm Sc}(r_p,\pi)$ does not include the parameter $\sigma_p$ and hence has one less free parameter than the streaming model. 

All equations above are based on the plane-parallel approximation. In the case of 6dFGS we also need to account for wide-angle effects. This means we will replace the equations above with more general descriptions, which we discuss in the next section.

\subsection{Wide-angle formalism}
\label{sec:wide}

So far we have assumed that the separation between galaxy pairs is much smaller than the distance of the galaxies from the observer. The 6dF Galaxy Survey has a maximum opening angle of $180^{\circ}$ and the (effective) redshift is fairly low at $z_{\rm eff} = 0.067$ (see section~\ref{sec:data}). We therefore include wide-angle correction terms. The wide-angle description of redshift-space distortions has been laid out in several papers~\citep{Szalay:1997cc,Szapudi:2004gh,Matsubara:2004fr,Papai:2008bd,Raccanelli:2010hk}. Here we will expand on this work by formulating the equations in terms of $g_{\theta}$ and $g_{b}$ and by distinguishing between the density-density, velocity-velocity and density-velocity contributions ($\xi_{\delta\delta}, \xi_{\theta\theta}$ and $\xi_{\delta\theta}$). The model will then correspond to the Scoccimarro model (eq.~\ref{eq:song}) in the last section.

The general redshift-space correlation function (ignoring the plane-parallel approximation) depends on $\phi$, $\theta$ and $s$. Here, $s$ is the separation between the galaxy pair, $\theta$ is the half opening angle, and $\phi$ is the angle of $s$ to the line-of-sight (see Figure~$1$ in~\citealt{Raccanelli:2010hk}). The angles $\phi$ and $\theta$ are not independent, but the relation between them is usually expressed through the two angles $\phi_1$ and $\phi_2$ given by $\phi = \frac{1}{2}(\phi_1 + \phi_2)$ and $\theta = \frac{1}{2}(\phi_1 - \phi_2)$. The total correlation function model, including $O(\theta^2)$ correction terms, is then given by~\citep{Papai:2008bd}
\begin{equation}
\begin{split}
\xi(\phi,\theta,s) &= a_{00} + 2a_{02}\cos(2\phi) + a_{22}\cos(2\phi) + b_{22}\sin^2(2\phi)\\
& +\Big[ - 4a_{02}\cos(2\phi) - 4a_{22} - 4b_{22} - 4a_{10}\cot^2(\phi)\\
& + 4a_{11}\cot^2(\phi) - 4a_{12}\cot^2(\phi)\cos(2\phi) + 4b_{11}\\
& - 8b_{12}\cos^2(\phi)\Big]\theta^2 + O(\theta^4).
\end{split} 
\label{eq:wide1}
\end{equation}
This equation reduces to the plane-parallel approximation if $\theta = 0$. The factors $a_{xy}$ and $b_{xy}$ in this equation are given by
\begin{equation}
\begin{split}
a_{00} &= g_b^2\xi_{0,\delta\delta}^2(r) + \frac{2g_bg_{\theta}}{3}\xi_{0,\delta\theta}^2(r) + \frac{2g_{\theta}^2}{15}\xi_{0,\theta\theta}^2(r)\\
& - \frac{g_bg_{\theta}}{3}\xi^2_{2,\delta\theta}(r) + \frac{2g_{\theta}^2}{21}\xi^2_{2,\theta\theta}(r) + \frac{3g_{\theta}^2}{140}\xi^2_{4,\theta\theta}(r)\\
a_{02} &= -\frac{g_bg_{\theta}}{2}\xi^2_{2,\delta\theta}(r) + \frac{3g_{\theta}^2}{14}\xi^2_{2,\theta\theta}(r) + \frac{g_{\theta}^2}{28}\xi^2_{4,\theta\theta}(r)\\
a_{22} &= \frac{g_{\theta}^2}{15}\xi^2_{0,\theta\theta}(r) - \frac{g_{\theta}^2}{21}\xi^2_{2,\theta\theta}(r) + \frac{19g_{\theta}^2}{140}\xi^2_{4,\theta\theta}(r)\\
b_{22} &= \frac{g_{\theta}^2}{15}\xi_{0,\theta\theta}^2(r) - \frac{g_{\theta}^2}{21}\xi^2_{2,\theta\theta}(r) - \frac{4g_{\theta}^2}{35}\xi^2_{4,\theta\theta}(r)\\
a_{10} &= \left[2g_bg_{\theta}\xi^1_{1,\delta\theta}(r) + \frac{4g_{\theta}^2}{5}\xi^1_{1,\theta\theta}(r)\right]\frac{1}{r} - \frac{g_{\theta}^2}{5r}\xi^1_{3,\theta\theta}(r)\\
a_{11} &= \frac{4g_{\theta}^2}{3r^2}\left[\xi^0_{0,\theta\theta}(r) - 2\xi^0_{2,\theta\theta}(r)\right]\\
a_{12} &= \frac{g_{\theta}^2}{5r}\left[2\xi^1_{1,\theta\theta}(r) -3\xi^1_{3,\theta\theta}(r)\right]\\
b_{11} &= \frac{4g_{\theta}^2}{3r^2}\left[\xi^0_{0,\theta\theta}(r) + \xi^0_{2,\theta\theta}(r)\right]\\
b_{12} &= \frac{2g_{\theta}^2}{5r}\left[\xi^1_{1,\theta\theta}(r) + \xi^1_{3,\theta\theta}(r)\right] ,
\end{split}
\label{eq:co}
\end{equation}
where the spherical harmonic moments, $\xi^m_{\ell,xy}(r)$, are
\begin{equation}
\begin{split}
\xi^m_{\ell,xy}(r) &= \int^{\infty}_{0}\int^1_{-1}\frac{k^mdkd\mu}{(2\pi)^2}e^{-(k\mu\sigma_v)^2}\cr
&\times \cos(kr\mu)Q_{xy}(k)\mathcal{P}_{\ell}(\mu),
\end{split}
\label{eq:xilm2}
\end{equation}
with $Q_{\delta\delta}(k)$, $Q_{\delta\theta}(k)$ and $Q_{\theta\theta}(k)$ as defined in equation~\ref{eq:D3} (see appendix~\ref{sec:ana} for an analytic solution).

In order to obtain a model for the 2D correlation function (including wide-angle effects), we can simply integrate eq.~\ref{eq:wide1} over $\theta$. In case of the plane-parallel approximation this would correspond to eq.~\ref{eq:song}. To reduce the equation to the simple streaming model we have to set $\sigma_v = 0$, $Q_{\delta\delta}(k) = Q_{\delta\theta}(k) = Q_{\theta\theta}(k)$ and convolve with $F(v)$.

\citet{Samushia:2011cs} studied wide angle effects in the SDSS-LRG sample and found that such effects are very small. The lower redshift and larger sky coverage of the 6dF Galaxy Survey mean that wide-angle effects are certainly larger in our data set than SDSS-LRG. However, we found no significant impact of such effects on our results. This is mainly because our analysis includes only rather small scales ($\pi < 30h^{-1}\,$Mpc, $r_p < 30h^{-1}\,$Mpc, see section~\ref{sec:fit}). This agrees with the findings of~\citet{Beutler:2011hx}.

We stress that the correction terms discussed above only capture first-order effects and so we additionally restrict our analysis to $\theta < 50^{\circ}$.

Many authors have found, using N-body simulations, that both the simple streaming and Scoccimarro models do not capture all non-linear and quasi-linear effects present in redshift surveys (e.g.~\citealt{Jennings:2010ne,Kwan:2011hr,Torre:2012} and references therein). For example the linear bias model seems to be too simplistic at small scales. In this study we restrict ourselves to the models discussed above but would like to point out that many other models have been suggested~\citep{Matsubara:2007wj,Matsubara:2008wx,Taruya:2010mx,Reid:2011ar,Seljak:2011tx} which could be compared to our data in future work.

\subsection{Systematics and the Alcock-Paczynski effect}
\label{sec:AP}

A dataset such as 6dFGS, containing galaxies with a high linear bias factor, may be prone to scale-dependent galaxy bias on small scales. We used the GiggleZ simulations~(Poole et al., in preparation) to derive the form of the scale-dependent bias. We find a $1.7\%$ correction to a linear bias at scales of $s = 10h^{-1}\,$Mpc, and note that this correction has a small but non-negligible impact on our results, depending on the smallest scales included in the fit.

A further systematic uncertainty comes from the so-called Alcock-Paczynski (AP) effect~\citep{Alcock:1979mp}. By assuming a cosmological model for the conversion of redshifts and angles to distances (see section~\ref{sec:data}) we could introduce an additional anisotropic signal in the correlation function which depends on the angle to the line-of-sight, if the true cosmology differs from our fiducial cosmological model. These distortions are degenerate with the linear redshift-space distortion signal and hence both effects need to be modelled to avoid systematic bias~\citep{Ballinger:1996cd,Simpson:2009zj}. Using the two scaling factors 
\begin{align}
f_{\parallel} &= \frac{H^{\rm fid}(z)}{H(z)}\\
f_{\perp} &= \frac{D_A(z)}{D^{\rm fid}_A(z)},
\end{align}
we can account for the AP effect by rescaling the corresponding axis
\begin{align}
\pi' &= \frac{\pi}{f_{\parallel}}\\
r_p' &= \frac{r_p}{f_{\perp}}.
\end{align}
While this is a matter of concern for high-redshift surveys, the 6dF Galaxy Survey is very nearly independent of the Alcock-Paczynski distortions because the distances $\pi$ and $r_p$ are almost independent of the fiducial cosmological model when expressed in units of $h^{-1}$Mpc. For example the value of $f_{\perp}$ at $z = 0.067$ changes by only ($0.3\%$, $1.5\%$) for $10\%$ changes in ($\Omega_m$, $w$). The corresponding values for $f_{\parallel}$ are ($0.7\%$, $1.6\%$). The relative tangential/ radial distortion depends on the value of $(1+z)D_A(z)H(z)/c$ relative to the fiducial model:
\begin{equation}
\frac{D_A(z)H(z)}{D_A^{\rm fid}(z)H^{\rm fid}(z)} = \frac{f_{\perp}}{f_{\parallel}}.
\end{equation}
This parameter combination changes by only $0.3$ ($0.1$)\% for $10\%$ changes in $\Omega_m$ ($w$).

Finally, it has been shown that $\gamma$ has a degeneracy with the equation of state parameter for dark energy, $w$~\citep{Simpson:2009zj}. High redshift measurements usually assume $w=-1$, which could introduce a bias in the measured value of $\gamma$ if dark energy is not exactly a cosmological constant. Again 6dFGS is robust against such effects.

\section{Fitting the 2D correlation function}
\label{sec:fit}

\begin{table}
\begin{center}
\caption{Cosmological parameters derived from the 6dFGS 2D correlation function. The effective redshift is $z_{\rm eff} = 0.067$. The last column indicates the priors/assumptions which go into each individual parameter measurement. The prior on the Hubble constant comes from~{\protect \citet{Riess:2011yx}} and the WMAP7 prior from~{\protect \citet{Komatsu:2010fb}}. The asterisks denote parameters which are derived from fitting parameters.}
\begin{tabular}{lll}
\hline
\multicolumn{3}{c}{Summary of parameter measurements from 6dFGS}\\
\hline
$g_{\theta}(z_{\rm eff})$ & $0.423\pm0.055$ & \\
$g_b(z_{\rm eff})$ & $1.134\pm0.073$ & \\
$\beta^*$ & $0.373\pm 0.054$ & \\
\hline
$\sigma_8$ & $0.76\pm 0.11$ & $\left[H_0=73.8\pm2.4, \gamma = 0.55\right]$\\
$\Omega_m$ & $0.250\pm0.022$ & $\left[H_0=73.8\pm2.4, \gamma = 0.55\right]$\\
b & $1.48\pm0.27$ & $\left[H_0=73.8\pm2.4, \gamma = 0.55\right]$\\
$f^*(z_{\rm eff})$ & $0.58\pm0.11$ & $\left[g_{\theta}+\sigma_8 \text{ from 6dFGS}\right]$\\
\hline
$\gamma$ & $0.547\pm 0.088$ & $\left[ \rm WMAP7\right]$\\
$\Omega_m$ & $0.271\pm0.027$ & $\left[ \rm WMAP7\right]$\\
\hline
\end{tabular}
\label{tab:para}
\end{center}
\end{table}

We now fit the two models for the 2D correlation function we developed earlier ($\xi_{\rm st}(r_p,\pi)$ and $\xi_{\rm Sc}(r_p,\pi)$) to our data. For the final constraints on $f\sigma_8$ we use the $\xi_{\rm Sc}(r_p,\pi)$ model since it gives similar results to the streaming model with one less free parameter.

Other studies (e.g.~\citealt{Samushia:2011cs}) prefer to analyse the correlation function moments $\xi_0, \xi_2$ (and if possible $\xi_4$), which carry the same information as the 2D correlation function. The correlation function moments have the advantage that the number of bins grows linearly with the highest scales analysed, while for $\xi(r_p,\pi)$ the number of bins grows quadratically. This makes it easier to get reliable covariance matrices. \citet{Samushia:2011cs} also show that the measurement errors of $\xi_{\ell}$ are more Gaussian. However, the correlation function moments are integrals over $\mu$ and hence carry information from all directions, including $\mu = 0$. Finger-of-God distortions can influence the correlation function moments up to large scales ($20-30h^{-1}\,$Mpc), while in the 2D correlation function they can be excluded via a cut in $r_p$. In 6dFGS we have found that these non-linear effects have a strong impact on the correlation function moments up to $30h^{-1}\,$Mpc. We have therefore decided to focus on the 2D correlation function instead of the correlation function moments.

\subsection{Derivation of the growth rate, $g_{\theta} = f\sigma_8$}

\begin{figure}
\begin{center}
   \epsfig{file=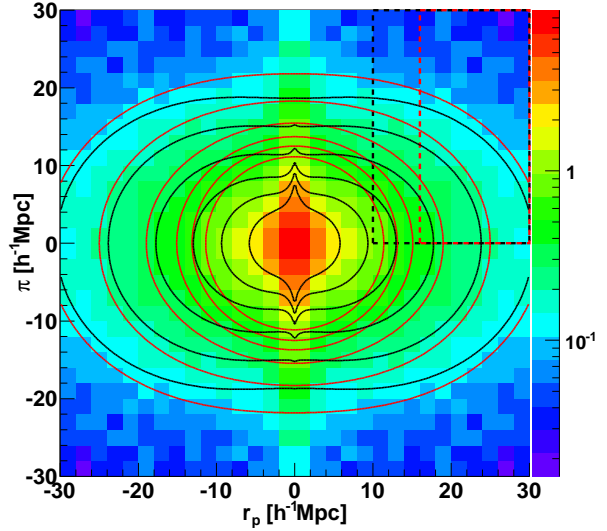,width=8cm}
\caption{The 2D correlation function in $2h^{-1}\,$Mpc bins. The fitting area is indicated by the dashed lines, where black corresponds to the streaming model, ($\xi_{\rm st}(r_p,\pi)$) and red corresponds to the Scoccimarro model ($\xi_{\rm Sc}(r_p,\pi)$). The black and red contours show the best fitting models for $\xi_{\rm st}(r_p,\pi)$ (black) and $\xi_{\rm Sc}(r_p,\pi)$ (red). The deviations seen in the two contours at large scales are well within the error bars of the two models, which can be seen in Figure~\ref{fig:chi2_1}. At small scales ($< 14h^{-1}\,$Mpc) the Scoccimarro model predicts much more clustering, while in the real data this clustering is smeared out along the line of sight because of the finger-of-God effect.}
   \label{fig:kaiser_models}
\end{center}
\end{figure}

\begin{figure}
\begin{center}
   \epsfig{file=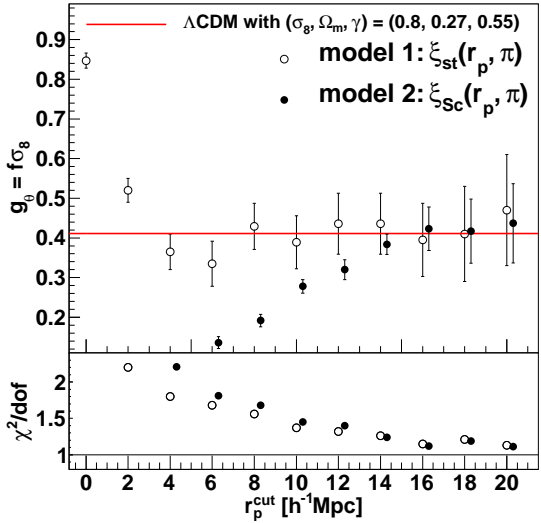,width=8cm}
\caption{The value of $g_{\theta}(z_{\rm eff}) = f(z_{\rm eff})\sigma_8(z_{\rm eff})$ as a function of the cut-off scale $r_p^{\rm cut}$, obtained by fitting the 6dFGS 2D correlation function with two different models (as described in section~\ref{sec:model} and~\ref{sec:wide}). At large scales the two models converge to similar values, while on small scales the models deviate from each other because of the different descriptions of non-linear evolution. For the final parameter measurements in Table~{\protect \ref{tab:para}} we chose model 2, $\xi_{\rm Sc}(r_p,\pi)$, with a conservative cut-off scale of $r_p^{\rm cut} = 16h^{-1}\,$Mpc. In the lower panel we plot the reduced $\chi^2$ as an indicator of the quality of the fit.}
   \label{fig:can}
\end{center}
\end{figure}

In Figure~\ref{fig:kaiser_models} we show the 6dFGS 2D correlation function. For our analysis we bin the data in $2\times 2h^{-1}\,$Mpc bins from $0$ to $30h^{-1}\,$Mpc in $r_p$ and $\pi$. Including larger scales does not add further information. At small $r_p$, the finger-of-God effect becomes dominant and we expect any linear model to fail. Since our description of non-linearities, in both of our models, is limited in its capability to capture all non-linear effects, it is necessary to include a cut-off scale $r^{\rm cut}_p$ marking a lower limit of the fitting range in $r_p$. 

Figure~\ref{fig:can} shows the measured value of $g_{\theta}$ as a function of the cut-off scale $r_p^{\rm cut}$ for our two different models. Above $r_p^{\rm cut} \approx 8h^{-1}\,$Mpc the streaming model, $\xi_{\rm st}(r_p,\pi)$, approaches a constant value of $g_{\theta}$. Our second model, $\xi_{\rm Sc}(r_p,\pi)$, contains a systematic error up to much larger scales, before it comes into agreement with the streaming model at about $r_p^{\rm cut} = 16h^{-1}\,$Mpc. This is expected since this model does not include a description of effects in the non-linear regime. For the final constraints we choose $r_p^{\rm cut} = 10h^{-1}\,$Mpc for the streaming model and $r_p^{\rm cut} = 16h^{-1}\,$Mpc for the Scoccimarro model. We also note that since the Scoccimarro model is based on only two free parameters ($g_{\theta}$ and $g_{b}$), the error is generally smaller compared to the streaming model, which has three free parameters ($g_{\theta}$, $g_b$ and $\sigma_p$). Other studies fit for the parameter $\sigma_v$ (e.g.~\citealt{Torre:2012}) in the Scoccimarro model, but we derive it using eq.~\ref{eq:sigv}.

For $\xi_{\rm Sc}(r_p,\pi)$ we use the fitting range $0 < \pi < 30h^{-1}\,$Mpc and $16 < r_p < 30h^{-1}\,$Mpc, which results in a total of $105$ bins. The best-fitting results are $g_{\theta} = 0.423\pm 0.055$ and $g_b = 1.134\pm 0.073$, where the errors for each parameter are derived by marginalising over all other parameters. The $\chi^2$ of this fit is $115$ with $103$ degrees of freedom (d.o.f.), indicating a good fit to the data. 

\begin{figure}
\begin{center}
   \epsfig{file=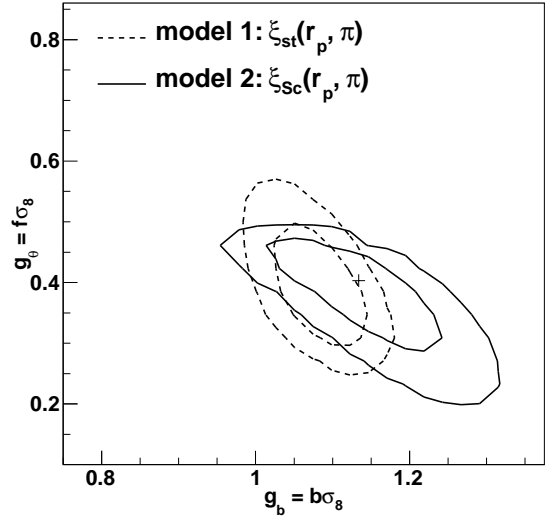,width=8cm}
\caption{Likelihood distribution of $g_{\theta}$ and $g_b$ derived from the fit to the 2D correlation function. The solid black contours show model $\xi_{\rm Sc}(r_p,\pi)$, while the dashed contours show the streaming model (see section~\ref{sec:model} and~\ref{sec:wide} for details of the modelling). The fitting range is $0 < \pi < 30h^{-1}\,$Mpc and $10 < r_p < 30h^{-1}\,$Mpc for $\xi_{\rm st}(r_p,\pi)$ and $0 < \pi < 30h^{-1}\,$Mpc and $16 < r_p < 30h^{-1}\,$Mpc for $\xi_{\rm Sc}(r_p,\pi)$. The black cross indicates the best-fitting value for the solid black contours.}
   \label{fig:chi2_1}
\end{center}
\end{figure}

For $\xi_{\rm st}(r_p,\pi)$, we have the fitting range $0 < \pi < 30h^{-1}\,$Mpc and $10 < r_p < 30h^{-1}\,$Mpc, which results in a total of $150$ bins. The best fitting parameters are $g_{\theta} = 0.389\pm 0.067$, $g_b = 1.084\pm 0.036$ and $\sigma_p = 198\pm81\,$km/s. The reduced $\chi^2$ of this fit is given by $\chi^2/\rm d.o.f. = 202/147 = 1.37$. We compare the constraints on $g_{\theta}$ and $g_b$ from both models in Figure~\ref{fig:chi2_1}.

In the Scoccimarro model we could use the parameter $\sigma_v \propto g_{\theta}$ instead of $g_{\theta}$ to test cosmology, as suggested by~\citet{Song:2010kq}. Our best fit gave $\sigma_v = 2.59\pm0.34h^{-1}\,$Mpc. However, this parameter depends on an additional integral over the velocity power spectrum, which adds a theoretical uncertainty. We therefore prefer to use $g_{\theta}$ in the following discussions. 

We can also express our results in terms of $\beta$ which is given by $\beta = g_{\theta}/g_b = 0.373\pm 0.054$. We summarise all measured and derived parameters in Table~\ref{tab:para}.

\subsection{Derivation of $\sigma_8$ and $\Omega_m$}
\label{sec:sig8}

\begin{figure}
\begin{center}
\epsfig{file=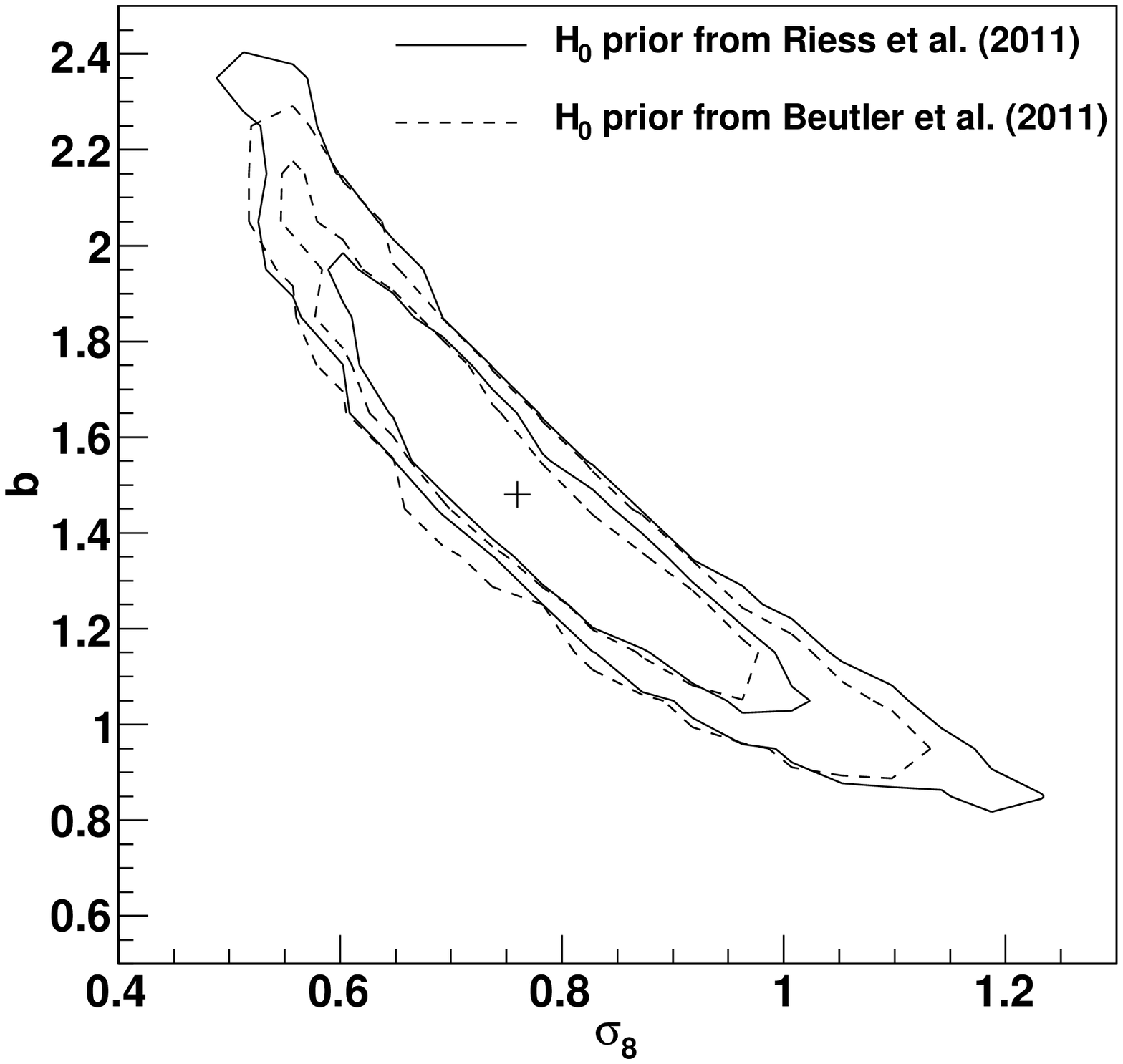,width=8cm}
\caption{This plot shows the likelihood distribution of the galaxy bias $b$ and $\sigma_8$, which we obtained by fitting the 6dFGS 2D correlation function assuming $\gamma = 0.55$. The solid black line shows the result using a prior on the Hubble constant of $H_0 = 73.8\pm2.4\,$km\,s$^{-1}$\,Mpc$^{-1}$ from~\citet{Riess:2011yx}, while the dashed black line uses a prior of $H_0 = 67\pm 3.2\,$km\,s$^{-1}$\,Mpc$^{-1}$ from~\citet{Beutler:2011hx}. Although the detection of redshift-space distortions can partially break the degeneracy between $b$ and $\sigma_8$ which exists in the 1D correlation function, there is still a significant residual degeneracy. The black cross marks the maximum likelihood value for the solid black lines.}
\label{fig:chi2_sigma8_b}
\end{center}
\end{figure}

In this section we use redshift-space distortions to directly measure $\sigma_8$. The angular dependence of the redshift-space distortion signal in the 2D correlation function allows us to measure $\beta$, which quantifies the amplitude of redshift-space distortions. Together with $\Omega_m(z)$ and $\gamma = 0.55$, this constrains the linear bias $b$ through the equation
\begin{equation}
b \simeq \frac{\Omega_m^{\gamma}(z)}{\beta}.
\end{equation}
Knowing $b$ we can use the absolute amplitude of the correlation function, $\left[b\sigma_8(z)\right]^2$, to constrain $\sigma_8(z{=}0) = [D(z{=}0)/D(z_{\rm eff})] \times \sigma_8(z_{\rm eff})$. 

For computational reasons we use our first model, $\xi_{\rm st}(r_p,\pi)$, in this sub-section and fit the five parameters $\sigma_8$, $\Omega_m$, $b$, $H_0$ and $\sigma_p$ using an MCMC approach. Since the shape of the correlation function is only sensitive to $\Gamma = \Omega_mh$, we cannot constrain $\Omega_m$ and $H_0$ at the same time. For the final results we include a prior on the Hubble constant ($H_0 = 73.8\pm2.4\,$km\,s$^{-1}$\,Mpc$^{-1}$,~\citealt{Riess:2011yx}, from now on referred to as HST prior) and marginalise over it. We use the same binning and fitting ranges as in the previous section.

The best-fitting model results in $\chi^2/\rm d.o.f = 1.35$. We find $\sigma_8 = 0.76\pm0.11$, $\Omega_m = 0.250\pm 0.022$, $b = 1.48\pm0.27$ and $\sigma_p = 174\pm 73\,$km/s. The remaining degeneracy between the bias $b$ and $\sigma_8$ is illustrated in Figure~\ref{fig:chi2_sigma8_b}. We include all these results in Table~\ref{tab:para}.

\begin{figure}
\begin{center}
\epsfig{file=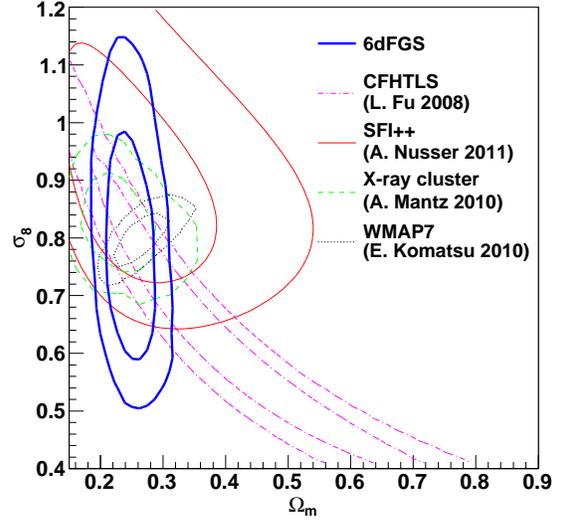,width=8cm}
\caption{This plot shows the likelihood distribution in $\sigma_8$ and $\Omega_m$ for different cosmological probes. The solid blue contours show the 6dFGS result, the magenta dotted dashed contours shows the currently best result of weak lensing from the CFHT wide synoptic Legacy Survey~\citep{Fu:2007qq}, the red solid contours show the result of the SFI++ peculiar velocity survey~\citep{Nusser:2011tu}, the green dashed contours show the result of~\citet{Mantz:2009fw} using cluster abundances and the black dotted contours are from WMAP7~\citep{Komatsu:2010fb}.}
\label{fig:chi2_om_sigma8}
\end{center}
\end{figure}

\begin{figure}
\begin{center}
\epsfig{file=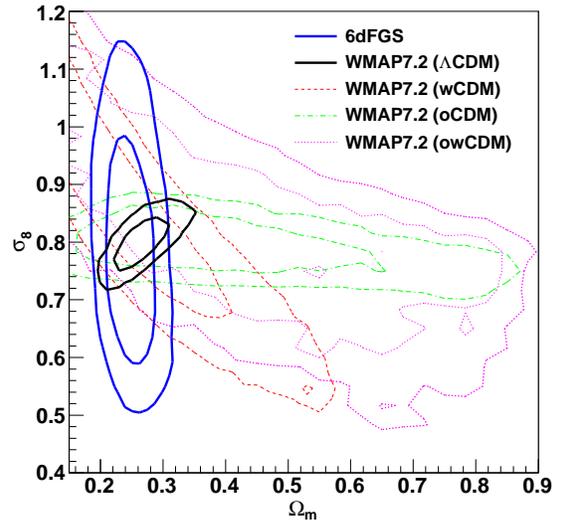,width=8cm}
\caption{Likelihood distribution of $\sigma_8-\Omega_m$ from WMAP7 for a $\Lambda$CDM model (solid black line), $w$CDM model (dashed red line), oCDM model (dotted dashed green line) and o$w$CDM model (dotted magenta line). In blue we show the 6dFGS result. Both parameters are defined at redshift zero and WMAP constraints on these parameters depend on assumptions about the expansion history of the Universe. We use CosmoMC~\citep{Lewis:2002ah}, together with the WMAP7.2~\citep{Komatsu:2010fb} dataset to produce these likelihood distributions.}
\label{fig:WMAP_om_sigma8}
\end{center}
\end{figure}

Figure~\ref{fig:chi2_om_sigma8} compares the 6dFGS $\Omega_m-\sigma_8$ probability distribution to measurements from several other datasets: The CFHT wide synoptic Legacy Survey (CFHTLS)~\citep{Fu:2007qq}, the SFI++ peculiar velocity survey~\citep{Nusser:2011tu}, cluster abundance from X-ray surveys~\citep{Mantz:2009fw} and WMAP7~\citep{Komatsu:2010fb}. Many of the experiments shown in this figure have systematic modelling uncertainties when extrapolating to $z = 0$ arising from assumptions about the expansion history of the Universe. Only 6dFGS and SFI++ are at sufficiently low redshift to be independent of such effects. To illustrate the impact of these effects we plot in Figure~\ref{fig:WMAP_om_sigma8} the probability distribution $\Omega_m-\sigma_8$ from WMAP7 for different cosmological models. The CMB measures the scalar amplitude $A_s$, which needs to be extrapolated from redshift $z_* \approx 1100$ to redshift zero to obtain $\sigma_8$. Every parameter that influences the expansion history of the Universe in this period affects the value of $\sigma_8$ derived from the CMB alone. The relation between $A_s$ and $\sigma_8$ is given by (e.g.~\citealt{Takada:2005si})
\begin{align}
\sigma_8^2(z) &= A_s\left(\frac{2c^2}{5\Omega_m H_0^2}\right)^2 \int^{\infty}_{0}k^3dkD^2(k,z)T^2(k)\left(\frac{k}{k_{*}}\right)^{n_s-1}\cr
&\times \left[\frac{3\sin(kR)}{(kR)^3} - \frac{3\cos(kR)}{(kR)^2}\right]^2,
\label{eq:sig8z}
\end{align}
where $R = 8h^{-1}\,$Mpc, $k_* = 0.02\,$Mpc$^{-1}$ and $A_s = (2.21\pm0.09) \times 10^{-9}$~\citep{Komatsu:2008hk}. $D(k,z)$ is the growth factor at redshift $z$ and $T(k)$ is the transfer function. If higher-order cosmological parameters such as the dark energy equation of state parameter $w$ are marginalised over, then the measurements of $\Omega_m$ and $\sigma_8$ weaken considerably (see Figure~\ref{fig:WMAP_om_sigma8}).

We now assess the influence of the $H_0$ prior. We replace the result of~\citet{Riess:2011yx} with a measurement derived from the 6dFGS dataset using Baryon Acoustic Oscillations~\citep{Beutler:2011hx}. The prior from this study is lower than the former value and is given by $H_0 = 67\pm 3.2\,$km\,s$^{-1}$\,Mpc$^{-1}$. Using the 6dFGS value of $H_0$ results in $\sigma_8 = 0.75\pm0.13$, $\Omega_m = 0.279\pm 0.028$, $b = 1.52\pm0.29$ and $\sigma_p = 174\pm 106\,$km/s. The quality of the fit is $\chi^2/\rm d.o.f. = 1.35$, very similar to the value obtained with the HST prior. Comparing the two results shows that a different prior in $H_0$ shifts the constraint in $\sigma_8$ and $b$ along the degeneracy shown in Figure~\ref{fig:chi2_sigma8_b}. However, we note that the 6dFGS measurement of $H_0$ is derived from the same dataset as our present study and hence could be correlated with our measured growth rate. We use these results only for comparison, and include the values obtained using the HST prior in Table~\ref{tab:para} as our final results of this section.

Alternative methods for deriving $\sigma_8$ or $f\sigma_8$ at low redshift are provided by peculiar velocity surveys (e.g.~\citealt{Gordon:2007zw,Abate:2009kd,Nusser:2011tu,Turnbull:2011ty,Davis:2010sw,Hudson:2012gt}). 6dFGS will soon provide its own peculiar velocity survey of around $10\,000$ galaxies. Velocity surveys have the advantage of tracing the matter density field directly, without the complication of a galaxy bias. However, they are much harder to obtain and current velocity surveys are $1-2$ orders of magnitude smaller than galaxy redshift surveys. 

\section{Cosmological implications}
\label{sec:impl}

\begin{figure}
\begin{center}
\epsfig{file=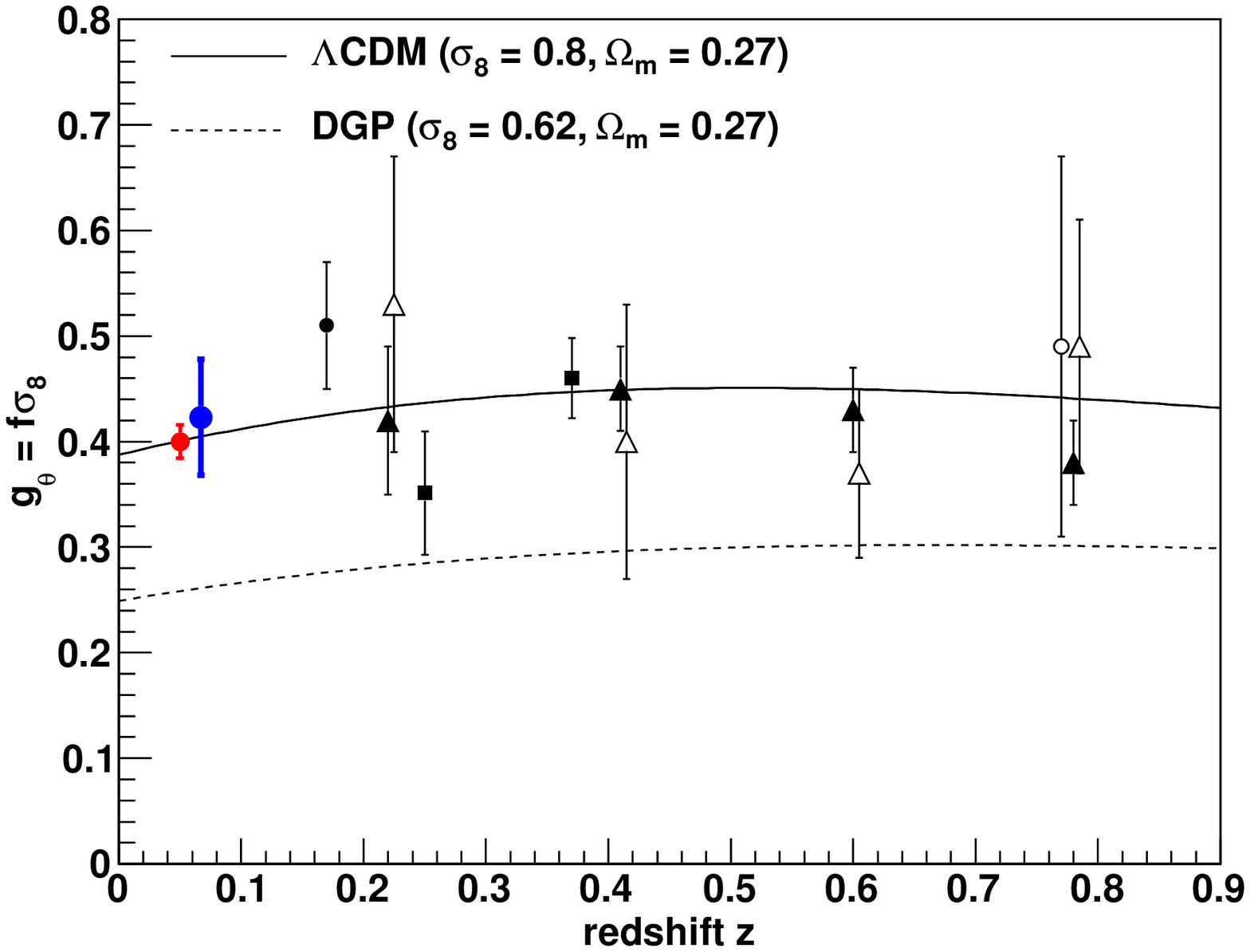,width=8cm}
\caption{Comparison of measurements of the growth of structure using galaxy surveys at different redshifts. The different data points belong to 6dFGS (solid blue circle, this paper), 2dFGRS (solid black circle;~\citealt{Hawkins:2002sg}), SDSS (solid black boxes;~\citealt{Samushia:2011cs}), WiggleZ (solid black triangles;~\citealt{Blake:2011rj}) and VVDS (empty circle;~\citealt{Guzzo:2008ac}). We also included a WALLABY forecast with a $4\%$ error-bar in red (see section~\ref{sec:future}). For the WiggleZ survey we also include the data points from~\citet{Blake:2011ep} (empty triangles, shifted by $\Delta z=0.005$ to the right for visibility), where the Alcock-Paczynski effect has been taken into account. We plot a $\Lambda$CDM model as well as a DGP model for comparison.}
\label{fig:growth}
\end{center}
\end{figure}

In this section we test General Relativity by measuring the growth index $\gamma$. We would like to stress that the $\gamma$-parameterisation of modified gravity has its limitations and other more general parameterisations have been proposed (see e.g.~\citealt{Silvestri:2009hh,Bean:2010zq,Daniel:2010yt,Clifton:2011jh,Hojjati:2011ix, Baker:2011jy}). However, this is going beyond the scope of this paper.

We combine our result for $g_{\theta}(z_{\rm eff})$ with the latest results from WMAP7~\citep{Komatsu:2010fb}, where we use the WMAP7.2 dataset provided on the NASA webpage\footnote{\url{http://lambda.gsfc.nasa.gov/product/map/dr4/likelihood_get.cfm}}. While $\Lambda$CDM predicts $\gamma \approx 0.55$, alternative theories of gravity deviate from this value. One example of such an alternative model is the DGP braneworld model of~\citet{Dvali:2000hr}, in which our observable Universe is considered to be a brane embedded in a higher dimensional bulk space-time and the leakage of gravity force propagating into the bulk can lead to the current accelerated expansion of the Universe. Because of the missing dark energy component, this model predicts a larger growth index of $\gamma \approx 0.69$~\citep{Linder:2005in}.

In Figure~\ref{fig:growth} we compare measurements of the growth of structure $g_{\theta}$ from different galaxy redshift surveys. For the WiggleZ survey we include data points which assume a correct fiducial cosmology (solid black triangles) as well as data points which account for the Alcock-Paczynski effect (empty black triangles). The degeneracy between the Alcock-Paczynski effect and the linear redshift-space distortion signal increases the error by about a factor of two. As we showed in section~\ref{sec:AP}, the Alcock-Paczynski effect is very small in 6dFGS, which therefore yields a direct measurement of the redshift-space distortion signal.

All data points seem to be in good agreement with the $\Lambda$CDM model (black solid line), while the DGP model generally predicts smaller values of $g_{\theta}$. The value of $\sigma_8$ for the two different models has been derived from the CMB scalar amplitude $A_s$~\citep{Komatsu:2010fb}, where we use the corresponding Friedmann equation to calculate $\sigma_8$.

The analysis method we apply in this section is summarised in the following four points:
\begin{enumerate}
\item We produce a Monte Carlo Markov Chain (MCMC) with CosmoMC~\citep{Lewis:2002ah} for a $\Lambda$CDM universe by fitting the WMAP7 dataset. The CMB depends on dark energy through the distance of last scattering and the late-time Integrated Sachs-Wolfe (ISW) effect. We avoid the contributions of the ISW effect by limiting the WMAP7 dataset to multipole moments $\ell > 100$.
\item Now we importance-sample the CosmoMC chain by randomly choosing a value of $\gamma$ in the range $0 \leq \gamma \leq 1$ for each chain element. Since the value of $\sigma_8(z_{\rm eff})$ depends on $\gamma$ we have to recalculate this value for each chain element. First we derive the growth factor
\begin{equation}
D(a_{\rm eff}) = \exp\left[ -\int_{a_{\rm eff}}^1da'\,f(a')/a'\right],
\end{equation}
where $a_{\rm eff}$ is the scale factor at the effective redshift $a_{\rm eff} = 1/(1 + z_{\rm eff})$. In order to derive $\sigma_{8,\gamma}(z_{\rm eff})$ we have to extrapolate from the matter dominated region to the effective redshift,
\begin{equation}
\sigma_{8,\gamma}(z_{\rm eff}) = \frac{D_{\gamma}(z_{\rm eff})}{D(z_{hi})}\sigma_8(z_{hi}),
\end{equation} 
where we use $\sigma_8(z_{hi})$ from eq.~\ref{eq:sig8z} and $z_{hi} = 50$, well in the matter-dominated regime.
\item We now calculate the growth rate using $f_{\gamma}(z_{\rm eff}) \simeq \Omega_m^{\gamma}(z_{\rm eff})$ and construct $g_{\theta,\gamma}(z_{\rm eff}) = f_{\gamma}(z_{\rm eff})\sigma_{8,\gamma}(z_{\rm eff})$.
\item Finally we compare the model with $g_{\theta} = 0.423\pm 0.055$ from Table~\ref{tab:para} and combine the likelihood from this comparison with the WMAP7 likelihood. 
\end{enumerate}

\begin{figure}
\begin{center}
\epsfig{file=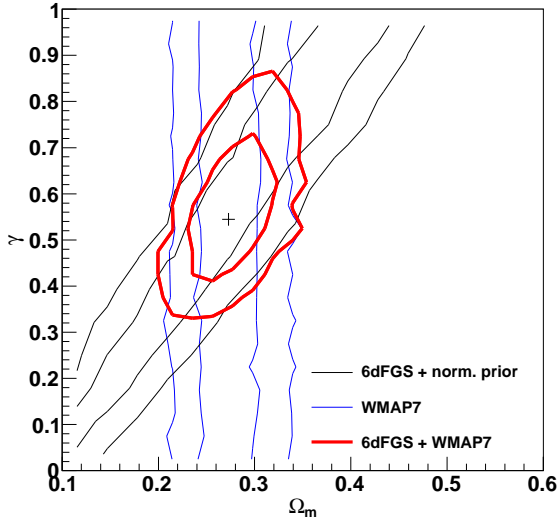,width=8cm}
\caption{Likelihood distribution of $\gamma-\Omega_m$ for our fit to $g_{\theta} = 0.423\pm0.055$ from 6dFGS and WMAP7~\citep{Komatsu:2010fb}. The 6dFGS contours (black) include a normalisation prior from WMAP7. Marginalising over the different parameters in the MCMC chain gives $\Omega_m = 0.271\pm 0.027$ and $\gamma = 0.547\pm 0.088$. The low redshift of 6dFGS makes this measurement particularly sensitive to $\gamma$ and independent of systematic effects like the Alcock-Paczynski distortion.}
\label{fig:gamma_om}
\end{center}
\end{figure}

The result is shown in Figure~\ref{fig:gamma_om}. Marginalising over the remaining parameters we get $\gamma = 0.547\pm 0.088$ and $\Omega_m = 0.271\pm 0.027$, which is in agreement with the prediction of a $\Lambda$CDM universe ($\gamma \approx 0.55$).  Our analysis depends only on the growth rate measured in 6dFGS and WMAP7. This makes our measurement of $\gamma$ independent of systematic effects like the Alcock-Paczynski distortion which is a matter of concern for galaxy redshift surveys at higher redshift. 

\section{Future low-redshift galaxy surveys: WALLABY and TAIPAN}
\label{sec:future}

In this section we make predictions for the accuracy of $f\sigma_8$ measurements from future low-redshift galaxy surveys using a Fisher matrix analysis based on~\citet{White:2008jy}

The Wide-field ASKAP L-band Legacy All-sky Blind surveY (WALLABY)\footnote{http://www.atnf.csiro.au/research/WALLABY} is an HI survey planned for the Australian SKA Pathfinder telescope (ASKAP), currently under construction at the Murchison Radio-astronomy Observatory (MRO) in Western Australia. The survey will cover $75\%$ of the sky and a proposal exists to fill up the remaining $25\%$ using the Westerbork Radio Telescope. In this analysis we follow the survey parameters employed by~\citet{Beutler:2011hx} (see also Duffy et al., in preparation), where for WALLABY we assume a $4\pi$ survey containing $600\,000$ galaxies at a mean redshift of $z = 0 .04$. The linear bias of a typical WALLABY galaxy is $0.7$~\citep{Basilakos:2007wq} and the volume of the survey is $0.12h^{-3}$Gpc$^3$.

The TAIPAN survey\footnote{TAIPAN: Transforming Astronomical Imaging surveys through Polychromatic Analysis of Nebulae} proposed for the UK Schmidt telescope at Siding Spring Observatory in New South Wales will cover a similar sky area as 6dFGS but will extend to a larger redshift such that $\overline{z} = 0.08$. For our TAIPAN forecast we assumed the same sky-coverage as 6dFGS ($f_{\rm sky} = 0.41$), a bias of $b = 1.4$, a total of $400\,000$ galaxies and a volume of $0.23h^{-3}$\,Gpc$^3$.

First we test the Fisher matrix prediction for $f\sigma_8$ in the case of 6dFGS. We assume a survey volume of $0.08h^{-3}\,$Gpc$^3$, with $81\,971$ galaxies. Using $k_{\rm max} = 0.1h\,$Mpc$^{-1}$ we forecast a measurement of $f\sigma_8$ of $23\%$, while using $k_{\rm max} = 0.2h\,$Mpc$^{-1}$ produces a $8.3\%$ error. The actual error in $f\sigma_8$ we found in this paper is $13\%$, somewhere between these two values. For WALLABY and TAIPAN we will report constraints for both $k_{\rm max} = 0.1h\,$Mpc$^{-1}$ and $k_{\rm max} = 0.2h\,$Mpc$^{-1}$.

With the specifications given above and using $k_{\rm max} = 0.1\; (0.2)h\,$Mpc$^{-1}$, the Fisher matrix forecast for WALLABY is a measurement of $f\sigma_8$ with $10.5\; (3.9)\%$ error. We included the WALLABY forecast with a $4\%$ error-bar in Figure~\ref{fig:growth}. The model TAIPAN survey produces forecast errors of $13.2\;(4.9)\%$, improving the results from 6dFGS by almost a factor of two. Although TAIPAN maps a larger volume of the Universe compared to WALLABY, it does not produce a better measurement of $f\sigma_8$. WALLABY has a smaller galaxy bias, which increases the redshift-space distortion signal by a factor of two compared to TAIPAN. This will also be very useful for breaking the degeneracy between bias and $\sigma_8$ using the technique of section~\ref{sec:sig8}.

The WALLABY survey will target galaxies rich in HI gas. Such galaxies will mostly populate under-dense regions of the Universe, because in groups and clusters galaxies are stripped of their gas by interactions with other galaxies and the intra-group and intra-cluster medium.  This is the reason that HI-selected galaxies possess a low bias ($\sim 0.7$).  However, this fact also implies that WALLABY galaxies sample the density field in a manner that avoids high-density regions. These high-density regions are an important source of non-linear redshift-space distortions (``finger-of-God'' effect). We can hence suppose that non-linear effects will be smaller in amplitude in an HI survey compared to highly-biased surveys such as 6dFGS (see e.g.~\citealt{Simpson:2011vn} or Figure 4 in~\citealt{Reid:2011ar}). This should allow the inclusion of much smaller scales in the analysis, producing more accurate measurements. A more detailed analysis using WALLABY mock catalogues is in preparation.

\citet{McDonald:2008sh} have suggested that multiple tracers within the same cosmic volume can be used to reduce the sampling variance and improve cosmological parameter constraints (see also \citealt{Seljak:2008xr,Slosar:2009aa,Bernstein:2011ju}). Using the ratio of the perturbation amplitudes of two surveys with different bias factors gives
\begin{equation}
\frac{b_1 + f\mu^2}{b_2 + f\mu^2} = \frac{\alpha b_2 + f\mu^2}{b_2 + f\mu^2},
\end{equation} 
where $\alpha = b_1/b_2$. The angular dependence of this expression allows one to extract $f$ without any dependence on the density field (see Figure 1 in \citealt{Bernstein:2011ju}). The density field is the source of the sampling variance error, since it will change, depending on the patch of the sky which is observed. Using the ratio of two tracers, the precision with which the growth rate $f$ can be determined is (in principle) only limited by the shot noise, and not by the sampling variance.

The three surveys discussed above, 6dFGS, WALLABY and TAIPAN, have a large overlapping volume which allows the potential application of this method. The technique works best for densely-sampled surveys with very different bias factors, $b_1$ and $b_2$. While TAIPAN and 6dFGS have very similar bias, the bias of WALLABY will be much smaller.

We assume an overlap volume of $0.41\times 0.12h^{-3}\,$Gpc$^3 = 0.049h^{-3}\,$Gpc$^3$, where we multiply the sky coverage of 6dFGS and TAIPAN with the effective volume of WALLABY. For the different surveys we use the parameters as stated above. We forecast Fisher matrix constraints on $f\sigma_8$ of  $10.3\;(5)\%$ using $k_{\rm max} = 0.1\;(0.2)h\,$Mpc$^{-1}$. Because the overlap volume is only $\approx 1/3$ of the WALLABY volume, this result does not improve the measurement arising from WALLABY alone, especially considering that WALLABY itself may be able to include modes up to large $k_{\rm max}$ in the fitting process.

Our results show that future surveys such as WALLABY and TAIPAN will provide an accurate measurement of $f\sigma_8$ at low redshift, and will be able to complement future high-redshift surveys such as BOSS, which will have a similar accuracy for several data points over the higher redshift range $0.2-0.6$~\citep{Song:2008qt,White:2008jy,Reid:2011ar}.

\section{Conclusion}
\label{sec:conc}

In this paper we have measured the 2D correlation function of the 6dF Galaxy Survey. We derived a covariance matrix using jack-knife resampling as well as log-normal realisations and showed that both techniques give comparable results. We have modelled the 2D correlation function with a simple streaming model and a more advanced approach suggested by~\citet{Scoccimarro:2004tg} combined with the N-body calibrated results from~\citet{Jennings:2010uv}. We formulated these models in real-space including wide-angle corrections. For the final results on $f\sigma_8$ we chose the model by~\citet{Scoccimarro:2004tg}, although we found that both models gave consistent results at sufficiently large scales. 

We analysed the measurement in two different ways. First we fitted for the two parameters $g_{\theta}(z_{\rm eff}) = f(z_{\rm eff})\sigma_8(z_{\rm eff})$ and $g_b(z_{\rm eff}) = b\sigma_8(z_{\rm eff})$, where these constraints depend only on the 6dFGS data. Our second analysis method assumes a growth index from standard gravity ($\gamma \approx 0.55$) and fits for $\sigma_8$, $b$, $\Omega_m$, $H_0$ and $\sigma_p$, where we combine the 6dFGS measurement with a prior in the Hubble constant. All parameter measurements are summarised in Table~\ref{tab:para}. We can summarise the results as follows:
\begin{itemize}
\item Our first analysis method found $g_{\theta}(z_{\rm eff}) = f(z_{\rm eff})\sigma_8(z_{\rm eff}) = 0.423\pm0.055$ and $g_b(z_{\rm eff}) = b\sigma_8(z_{\rm eff}) = 1.134\pm0.073$, at an effective redshift of $z_{\rm eff} = 0.067$. The 6dFGS measurement of $g_{\theta}$, unlike high-redshift measurements, does not depend on assumptions about the expansion history of the Universe and the Alcock-Paczynski distortion.
\item In our second analysis method we used the angle dependence of redshift-space distortions in the 2D correlation function to break the degeneracy between the galaxy bias $b$ and the normalisation of the matter clustering statistic $\sigma_8$, assuming standard gravity. We found $\sigma_8 = 0.76\pm0.11$, $\Omega_m = 0.250\pm 0.022$, $b = 1.48\pm0.27$ and $\sigma_p = 174\pm73$km/s. This result uses a prior on $H_0$ from~\citet{Riess:2011yx}.
\item Combining our measurement of $g_{\theta}(z_{\rm eff})$ with WMAP7~\citep{Komatsu:2010fb} allows us to measure the growth index $\gamma$, directly testing General Relativity. We found $\gamma = 0.547\pm 0.088$ and $\Omega_m = 0.271\pm 0.027$, in agreement with the predictions of General Relativity ($\gamma \approx 0.55$). The 6dFGS measurement of this parameter is independent of possible degeneracies of $\gamma$ with other parameters which affect the correlation function at high redshift, such as the dark energy equation of state parameter $w$.
\item We used a Fisher matrix analysis to forecast the constraints on $f\sigma_8$ that would be obtained from two future low-redshift galaxy surveys, WALLABY and TAIPAN. We found that WALLABY will be able to measure $f\sigma_8$ to a forecast accuracy of $10.5\%$ for $k_{\rm max} = 0.1h\,$Mpc$^{-1}$ and $3.9 \%$ for $k_{\rm max}=0.2h\,$Mpc$^{-1}$. A combination of 6dFGS, TAIPAN and WALLABY, using the multiple-tracer method proposed by \citet{McDonald:2008sh}, will be able to constrain $f\sigma_8$ to $5-10.3\%$. These measurements would complement future large-volume surveys such as BOSS, which will measure the growth rate at much higher redshift ($z > 0.2$), and contribute to future precision tests of General Relativity on cosmic scales.
\end{itemize}

\section*{Acknowledgments}

The authors thank Alex Merson for providing the random mock generator. We thank Alexandra Abate, Liping Fu, Patrick Henry, Adi Nusser, Adam Mantz and Alexey Vikihlinin for providing their results for comparison. We are also grateful to David Parkinson for help with CosmoMC and Yong-Seon Song, Takahiko Matsubara, Martin Meyer, Morag Scrimgeour and David Rapetti for fruitful discussions.

F.B. is supported by the Australian Government through the International Postgraduate Research Scholarship (IPRS) and by scholarships from ICRAR and the AAO. Part of this work used the iVEC$@$UWA supercomputer facility.
GBP acknowledges support from two Australian Research Council (ARC) Discovery Projects (DP0772084 and DP1093738).
The 6dF Galaxy Survey was funded in part by an Australian Research Council Discovery Projects Grant (DP-0208876), administered by the Australian National University.

\setlength{\bibhang}{2em}
\setlength{\labelwidth}{0pt}

\appendix

\appendix

\newpage

\section{Partial analytical solution for the correlation function moments integral}
\label{sec:ana}

The double integrals in eqs.~\ref{eq:xilm1} and \ref{eq:xilm2} are difficult to solve numerically. Here we show an analytical solution for the integral over $\mu$ which allows for a faster numerical solution of the full integral. First we re-write eq.~\ref{eq:xilm2} as
\begin{equation}
\begin{split}
\xi^m_{\ell,xy}(r) &= \int^{\infty}_{0}\int^1_{-1}\frac{k^mdkd\mu}{(2\pi)^2}e^{-(k\mu\sigma_v)^2}\cr
&\times \cos(kr\mu)D_{xy}(k)\mathcal{P}_{\ell}(\mu)\cr
& =  \int^{\infty}_{0}\frac{k^mdk}{(2\pi)^2}Q_{xy}(k) \int^1_{-1}d\mu\; e^{ikr\mu-(k\mu\sigma_v)^2}\mathcal{P}_{\ell}(\mu),\notag
\end{split}
\end{equation}
where $i$ is the complex number. More generally the integral over $\mu$ can be written as
\begin{equation}
\begin{split}
F_{n}(\mu)=&\int^1_{-1}d\mu\; e^{ikr\mu-(k\mu\sigma_v)^2}\mu^n\cr
=&\left(-i\frac{\partial}{\partial (kr)}\right)^{n}\int^1_{-1}d\mu\; e^{ikr\mu-(k\mu\sigma_v)^2}\cr
=&\left(-\frac{\partial}{\partial (k^2\sigma_v^2)}\right)^{n/2}\int^1_{-1}d\mu\; e^{ikr\mu-(k\mu\sigma_v)^2}.
\end{split}
\end{equation}
The integral on the right can be solved analytically. If we set $a = kr$ and $b = (k\sigma_v)^2$ we obtain
\begin{equation}
\begin{split}
&\int^1_{-1}d\mu\; e^{ikr\mu-(k\mu\sigma_v)^2} \cr
=& \frac{i}{\sqrt{b}}e^{-ia-b}\left[\text{Daw}\left(\frac{a - 2ib}{2\sqrt{b}}\right) - e^{2ia}\text{Daw}\left(\frac{a + 2ib}{2\sqrt{b}}\right)\right].
\end{split}
\end{equation}
Taking the n-th derivatives of the real part of the term above gives $F_n(\mu)$. The Dawson integral $\text{Daw}(x)$ can be calculated using the imaginary error function $\text{erfi}(x)$:
\begin{equation}
\begin{split}
\text{Daw}(x) &= e^{-x^2}\int^x_0e^{y^2}dy\cr
&= \frac{\sqrt{\pi}}{2}e^{-x^2}\text{erfi}(x).
\end{split}
\end{equation}
We can than construct eq.~\ref{eq:xilm2} for the different correlation function moments:
\begin{equation}
\begin{split}
\xi^m_{\ell=0,xy}(r) &= \int^{\infty}_{0}\frac{k^mdk}{(2\pi)^2}Q_{xy}(k) F_0(\mu)\cr
\xi^m_{\ell=1,xy}(r) &= \int^{\infty}_{0}\frac{k^mdk}{(2\pi)^2}Q_{xy}(k) F_1(\mu)\cr
\xi^m_{\ell=2,xy}(r) &= \int^{\infty}_{0}\frac{k^mdk}{(2\pi)^2}Q_{xy}(k) \frac{1}{2}\left[3F_2(\mu) - F_0(\mu)\right]\cr
\xi^m_{\ell=3,xy}(r) &= \int^{\infty}_{0}\frac{k^mdk}{(2\pi)^2}Q_{xy}(k) \frac{1}{2}\left[5F_3(\mu) - 3F_1(\mu)\right]\cr
\xi^m_{\ell=4,xy}(r) &= \int^{\infty}_{0}\frac{k^mdk}{(2\pi)^2}Q_{xy}(k) \frac{1}{8}\left[35F_4(\mu) - 30F_2(\mu) + 3F_0(\mu)\right]\cr
\dots&
\end{split}
\end{equation}
and so on.

\label{lastpage}

\end{document}